\documentstyle{lamuphys}
\makeatletter
\let\chapter\hid@chapter
\makeatother
\newcommand{\nc}{\newcommand}
\nc{\be}{\begin{equation}}
\nc{\ee}{\end{equation}}
\nc{\bea}{\begin{eqnarray}}
\nc{\eea}{\end{eqnarray}}
 
\nc{\eqn}[1]{{(\ref{#1})}}
\nc{\cA}{{\cal A}}
\nc{\cB}{{\cal B}}
\nc{\cC}{{\cal C}}
\nc{\cD}{{\cal D}}
\nc{\cE}{{\cal E}}
\nc{\cF}{{\cal F}}
\nc{\cG}{{\cal G}}
\nc{\cH}{{\cal H}}
\nc{\cI}{{\cal I}}
\nc{\cJ}{{\cal J}}
\nc{\cK}{{\cal K}}
\nc{\cL}{{\cal L}}
\nc{\cM}{{\cal M}}
\nc{\cN}{{\cal N}}
\nc{\cO}{{\cal O}}
\nc{\cP}{{\cal P}}
\nc{\cQ}{{\cal Q}}
\nc{\cR}{{\cal R}}
\nc{\cS}{{\cal S}}
\nc{\cT}{{\cal T}}
\nc{\cU}{{\cal U}}
\nc{\cV}{{\cal V}}
\nc{\cW}{{\cal W}}
\nc{\cX}{{\cal X}}
\nc{\cY}{{\cal Y}}
\nc{\cZ}{{\cal Z}}
\nc{\tr}{{{\rm tr}\,}}

\begin{document}
\pagenumbering{arabic}
\titlerunning{The Gross-Neveu model and QCDs chiral
phase transition}
\title{
\hfill{\small HD-THEP-97-47$\;$}\break
The Gross-Neveu model and \\
QCDs chiral phase transition}

\author{Thomas\,Reisz
\thanks{Supported by a Heisenberg Fellowship}
}

\institute{Institut f\"ur Theoretische Physik, Universit\"at Heidelberg,
Philosophenweg 16, \\
D-69120 Heidelberg, Germany}

\maketitle

\begin{abstract}
Quantum chromodynamics has a rather complicated phase structure.
The finite temperature, chiral phase structure depends on the
number of flavours and to a large extent on the particular values
of the fermion masses.
For two massless flavours there is a true second order transition.
It has been argued that this transition belongs to the universality class
of the three-dimensional O(4) spin model.
The arguments have been questioned recently,
and the transition was claimed to be mean field behaved.

In this lecture we discuss this issue at the example of the
three-dimensional, parity symmetric Gross-Neveu model
at finite temperature,
with a large number $N$ of fermions.
At zero temperature
there is a phase where parity is spontaneously broken.
At finite temperature, this model has a parity restoring second order
transition. It reveals considerable similarity to the QCD chiral phase
transition. There are related questions here concerning the universality
class.
We solve this problem essentially by means of the following methods:
Large $N$ expansion,
dimensional reduction in the framework of quantum field theory,
and high order convergent series expansions about disordered lattice
systems.
\end{abstract}
%
%%%%%%%%%%%%%%%%%%%%%%%%%%%%%%%%%%%%%%%%%%%%%%%%%%%%%%%%%%%
%%%%%%%%%%%%%%%%%%%%%%%%%%%%%%%%%%%%%%%%%%%%%%%%%%%%%%%%%%%
\section{Introduction - QCD and the chiral phase transition}
%%%%%%%%%%%%%%%%%%%%%%%%%%%%%%%%%%%%%%%%%%%%%%%%%%%%%%%%%%%
%%%%%%%%%%%%%%%%%%%%%%%%%%%%%%%%%%%%%%%%%%%%%%%%%%%%%%%%%%%

Quantum field theories provide the appropriate framework for
the understanding and the quantitative description
of high energy physics.
In particular,
quantum chromodynamics - QCD - is supposed to be the theory of
strong interactions, the physics of quarks, mesons and hadrons.
As such a theory it has to describe a large variety of high energy phenomena.
For instance this concerns the confinement of quarks,
the observed mesonic and hadronic particle mass spectrum,
topological excitations.
This theory evolves an enormous complexity.
Computational techniques to be developed and used are 
rather complicated, both numerical and analytical ones.

There are a number of long-term projects running, with the aim
of a quantitative
understanding of particular properties of QCD from first principles.
For example, one of these projects is the so-called alpha-collaboration
(\cite{luescher1}).
One of the aims is to determine the flow of the (appropriately defined)
renormalized coupling
constant $g_R(L)$ well above the perturbative short distance scale,
that is on large spatial scale $L$.
A remarkable strategy of this collaboration is the
combination of the appropriate analytical and numerical
(Monte-Carlo) techniques.

Other long-term projects concern the nature of the chiral phase
transition in finite temperature QCD and the properties of
the high temperature phase.
The high temperature phase of QCD may be realized in heavy ion collisions,
and it was certainly realized in the early universe.
There are a large number of interesting physical questions related to
this phase.
For instance,
screening of heavy quarks in the QCD plasma phase
has been studied in detail and
screening masses have been extracted, by combining analytical
methods such as finite volume perturbation theory and dimensional
reduction in the framework of quantum field theory,
together with various Monte-Carlo methods.
Other, yet open questions concern
the (gauge-invariant) definition of a magnetic mass and its determination,
the ability to disentangle infrared and ultraviolet singularities and their removal
in the thermodynamic limit,
that is to find appropriate resummation techniques.

In the limit of vanishing current quark masses,
massless QCD has a chiral symmetry.
This global symmetry is spontaneously broken at zero temperature,
and it becomes restored in the high temperature phase.
Whether this symmetry is restored by a 1st or 2nd order phase transtion
or just by a crossover phenomenon depends on the number of 
fermionic flavours.
In addition, to a large extent it depends on the
particular values of the fermion masses.

A detailed presentation of the 
QCD phase transitions according to our current knowledge
can be found in the review of \cite{hildegard1}.
Here we only give a very short outline of the state of the art.
In Fig.~1 we show a qualitative plot of the
supposed phase diagram for three flavours $u,d,s$
in dependence on their masses.
The $u$- and $d$-quark masses are of comparable size and
are put to the same value.
For the pure SU(3) Yang-Mills theory (right upper corner)
the transition is the deconfinement transition, which is believed to be
of 1st order. For the
purely massless case (left lower corner) the transition refers to
the chiral transition, which is supposed to be of first order
in the case of 3 flavours.
With increasing masses the chiral transition becomes weaker
and finally turns over into a mere
crossover.
For 2 massless flavours, the chiral transition is of first order below
a critical value $m_s^*$ of the strange quark mass
(corresponding to a tricritical point)
and of second order above $m_s^*$.

The physical fermion masses are subject to a large uncertainty
in relation to the phase boundaries.
Even if there is no "true" phase transition in QCD,
some correlation lengths might become large, generating large domains
of chiral condensates.

%%%%%%%%%%%%%%%%%%%
% Fig. QCD phase diagram
%%%%%%%%%%%%%%%%%%%
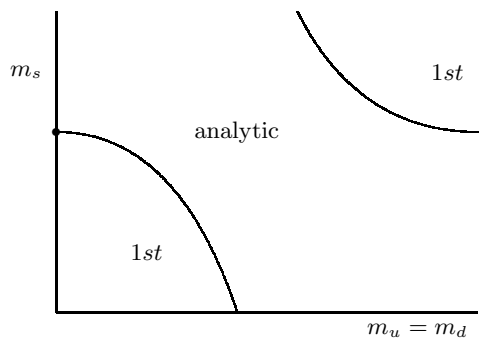
\begin{figure}

%\begin{center}
\setlength{\unitlength}{0.8cm}
\begin{picture}(10.0,5.5)
 
% diagram
 
\put(2.0,0.0){
\setlength{\unitlength}{0.8cm}
\begin{picture}(8.0,7.0)
 
% coordinate axis
\put(0.0,0.0){\line(1,0){7.0}}
\put(-1.0,4.0){\makebox(1.0,0){$m_s$}}
\put(0.0,0.0){\line(0,1){5.0}}
\put(5.5,-0.3){\makebox(1.0,0){$m_u=m_d$}}
 
% sectors
\qbezier(0.0,3.0)(2.0,3.0)(3.0,0.0)
\put(1.0,1.0){\makebox(1.0,0){$1st$}}
\qbezier(7.0,3.0)(5.0,3.0)(4.0,5.0)
\put(6.0,4.0){\makebox(1.0,0){$1st$}}
%\put(0.1,3.2){\makebox(0.3,0){$\times$}}
\put(2.5,3.0){\makebox(1.0,0){analytic}}
% critical and tricitical
\put(0.0,3.0){\circle*{0.13}}

\end{picture}
}
  
\end{picture}

%\end{center}
\caption{\label{qcdphase} A qualitative plot of the mass dependence
of the QCD phase
transition for three flavours.
The dot identifies the tricitical strange quark mass,
above which the transition is of 2nd order.
}

\end{figure}

%%%%%%%%%%%%%%%%%%%
% end Fig. QCD phase transition
%%%%%%%%%%%%%%%%%%%

In nature, two fermion flavours are realized 
as relatively light states.
For the case of two flavours, both being massless,
the chiral phase transition
is supposed to be described by a critical point.
What is the nature of this transition?
It was argued by Pisarski and Wilczek that its
universality class is that
of the O(4) spin model in 3 dimensions.
We will discuss this point of view in the next section.
On the contrary, Kocic and Kogut argued that the transition is mean
field behaved.
So far, no definite answer has been given, neither by Monte Carlo
simulations nor by analytical investigations.

In this lecture we discuss the issue of the finite temperature phase
transition at the example of the 3-dimensional Gross-Neveu model.
This model reveals considerable similarity to QCD.
There are quite related questions concerning the nature of
the transition, including as candidates both mean field and non-trivial
spin model behaviour.
However, in this case a definite answer can be given,
mainly by analytical means.
 
%%%%%%%%%%%%%%%%%%%%%%%%%%%%%%%%%%%%%%%%%%%%%%%%%%%%%%%%%%%
\subsection{Two-flavour QCD}
%%%%%%%%%%%%%%%%%%%%%%%%%%%%%%%%%%%%%%%%%%%%%%%%%%%%%%%%%%%

Before we turn over to the Gross-Neveu model, let us discuss
in some more detail the effective scalar O(4) model for the
QCD chiral phase transition.
The order parameter is given by the chiral condensate
$<\overline\psi(x) \psi(x)>$.
It is argued that the transition belongs to the
universality class of the scalar O(4) model in 3 dimensions
(\cite{pisarski}, \cite{wilczek}, \cite{rajagopal}).
An essential ingredient of this argument
is the observation that under quite
general conditions, for a field theory
at high temperature
fermionic degrees of freedom act only as composite states
in the infrared. This fact is a consequence of the so-called
dimensional reduction in quantum field theory and
will be discussed in detail in the following
sections.

Let us outline the way of reasoning towards the O(4) scenario.
The idea is to build an effective model of the "localized" order
parameter field, the condensate,
and apply a Landau-Ginzburg type of argument.
First of all, the order parameter field is a complex 
$2\times 2$-matrix field in three dimensions $x\in{\bf R}^3$,
\be
   < \overline\psi_{Ri}(x) \psi_{Lj}(x) > \sim M_{ij}(x)
   \in M(2\times 2, {\bf C}),
\ee
where $R$ and $L$ denotes right- and left-handed (relativistic)
fermions, respectively.
Close to the critical point the degrees of freedom that develop
long-range fluctuations are assumed to be associated
with the order parameter $M$, which is small in magnitude.
Hence it is plausible to describe the effective model by the
action
\bea
   S_{eff}(M) = \int_x && ( \tr \partial_x M(x)^\dagger \partial_x M(x)
   + m^2 \; \tr M^\dagger(x) M(x) \\
   && + g_1 \tr (M^\dagger(x) M(x) )^2
         + g_2 (\tr M^\dagger(x) M(x) )^2 ) . \nonumber
\eea
The partition function becomes
\be
   Z = \int \cD M \; \cdot \exp{(-S_{eff}(M))},
\ee
where $\cD M = \prod_x dM(x)$.
We have implicitly assumed an ultraviolet cutoff, such as the
Pauli-Villar cutoff or a lattice.
At this stage then we can exploit the methods of renormalizable
quantum field theory, such as the renormalization group and
renormalized perturbation theory, and strong statements
obtained by those methods.
On the one hand, a posteriori this justifies a large
freedom in the choice of the ultraviolet cutoff for the above
(super-) renormalizable model.
That is, a particular choice does not have influence
on the low-momentum scale.
On the other hand, the critical properties of the model can be studied.

Yet the model has too large a symmetry to describe a chiral phase
transition. It is invariant under global $U(2)_L \times U(2)_R$
transformations
\be
     M \; \to \; U M V^\dagger \; , \quad U,V\in U(2).
\ee
Finally we want to reduce the symmetry to the smaller,
physical symmetry group
$SU(2)_L \times SU(2)_R \times U(1)$,
(to be broken down to vector $SU(2) \times U(1)$ spontaneously
at low temperature,)
where $U(1)$ describes the vector-baryon symmetry.
This restricted symmetry has to be implemented by imposing
appropriate constraints on $M$. 
Whereas for larger number of flavours those constraints are
to be non-linear in $M$, leaving the framework of renormalizable
quantum field theory,
for two flavours we can implement such  constraints by
restricting $M$ to be a real multiple of a unitary matrix of
unit determinant,
$M \in c \cdot SU(2) \; , \; c\in{\bf R}$.
Hence $M$ can be parametrized linearly as
\be
   M \; = \; \pi_0 {\bf 1}_2 + \vec\pi \vec\sigma \; ,
    \; (\pi_0,\vec\pi)\in{\bf R}^4,
\ee
where $\vec\sigma = (\sigma_1,\sigma_2,\sigma_3)$ denotes the
Pauli-matrices.
It is now obvious that in terms of the $\pi$-fields we get a linear
O(4) model. This symmetry is broken at low temperature
to O(3).

If the universality arguments made so far are true, the QCD chiral
phase transition for two massless flavours will belong to the
universality class of the O(4) spin model in three dimensions.
Of course, there are assumptions made in the course 
towards this conclusion.
The most stringent ones are the following.

\begin{itemize}
\item The chiral phase transition is of 2nd order.
\item The free energy density as a function of the order
      parameter field $M$ is sufficiently smooth, allowing for
      a polynomial representation at small $M$.
      But we know this is problematic below the upper critical dimension,
      the dimension of strict renormalizability,
      which is 4 for the above model.
\item In the language of statistical mechanics, the quantum critical
     system is equivalent to the corresponding classical critical system.
     In particular, the phase transition temperature is so high that
     the thermal fluctuations dominate the quantum fluctuations
     even at the transition.
     In the language of quantum field theory, this amounts to say
     that dimensional reduction is complete at the critical point,
     at least concerning the values of the critical exponents.
\end{itemize}

Recently Kocic and Kogut
have criticised the above ($\sigma$-model) approach because the
fermionic degrees of freedom are represented only via composite
fields $M$
(\cite{kk1}).
Actually, as we will see in the course of this lecture, this is a
minor point of criticism.
But even if fermions act only as composite states,
there are several candidates for the universality class of the
QCD chiral phase transition.
Both the above argumentations mainly refer to conventional wisdom.
In order to obtain a definite answer it is unavoidable to give
strong reasonings,
including explicit computations.

%%%%%%%%%%%%%%%%%%%%%%%%%%%%%%%%%%%%%%%%%%%%%%%%%%%%%%%%%%%
\subsection{Why studying the Gross-Neveu model?}
%%%%%%%%%%%%%%%%%%%%%%%%%%%%%%%%%%%%%%%%%%%%%%%%%%%%%%%%%%%
In this lecture the problem of the chiral phase transition will be
discussed at the example of a much simpler model than QCD, which
nevertheless bears the essential properties such as "chirality".
Namely, I would like to discuss the 3-dimensional Gross-Neveu model
at finite temperature, with a large number of fermionic flavours.
It is a model with a four-fermion interaction, subject to a global
${\bf Z}_2$ symmetry (which is parity).
This symmetry is known to be broken at zero temperature, and it
is restored at high temperature by a 2nd order transition.
We then ask for the universality class of this transition.

In quite analogy to QCD there are two opposite claims
of the universality class. Adjusting the arguments of
Pisarski and Wilczek, the transition is described by a
${\bf Z}_2$ symmetric scalar model in 2 dimensions that
should belong to the Ising class.
On the contrary, Kocic and Kogut claimed that the
transition is that of mean field theory.

Beyond its similarity to QCD, the advantage of studying this model is that
to a large extent we have a closed analytic control of it.
This includes the finite temperature transition.
As we will see, the above discrepancy can be resolved
by analytical means.
Also, in the course of the following sections it will become
clear that the definite answer can be given only by
explicit computations.

The answer will be given essentially by applying the following two
techniques.
\begin{itemize}
\item Dimensional reduction in quantum field theory.
Applied to the 3-di\-men\-sional, finite temperature Gross-Neveu model,
it states that
the infrared behaviour of the latter is described by a 2-dimensional
scalar field theory at zero temperature. As we will see,
this effective model is local and renormalizable.
It should be emphasized that dimensional reduction provides an
explicit computational device to compute the couplings of
the effective model.
This is important because a 2-dimensional scalar model does not
necessarily belong to the Ising universality class.
\item Convergent series expansions of high orders, applied
to the free energy and connected correlations.
In this way the critical region can be investigated, the order
of the transition and eventually the universality class can
be determined.
\end{itemize}

In outline, this lecture is organized as follows.
Throughout we will use the path integral representation for
quantum field theories.
In Sect. 2 we describe the 3-dimensional Gross-Neveu model.
The notions familiar from finite temperature quantum field theory
are introduced. The large $N$ representation and expansion of
the model are discussed.
The $N=\infty$ limit can be solved in closed form, including the
phase structure.
In Sect. 3 we outline the technique of
dimensional reduction in the framework of renormalizable
quantum field theory.
In the recent past this technique became a very
well developed, powerful computational machinery both for
bosonic and fermionic field theories at finite temperature.
It is applied to the Gross-Neveu model.
Then we are prepared to study in Sect. 4 the phase structure
of the model for finite number $N$ of flavours.
A first insight is obtained by the
discussion of the strong coupling limit.
In a subsection we shortly present convergent series expansion
techniques, in particular linked cluster expansions, and the ideas
behind their generation.
They are then applied to determine the critical exponents,
hence the universality class of the parity restoring transition.
Summary and outlook for the case of QCD is given in the
last Sect. 5.

%%%%%%%%%%%%%%%%%%%%%%%%%%%%%%%%%%%%%%%%%%%%%%%%%%%%%%%%%%%
%%%%%%%%%%%%%%%%%%%%%%%%%%%%%%%%%%%%%%%%%%%%%%%%%%%%%%%%%%%
\section{\label{gn}
The 3-dimensional Gross-Neveu model at zero and finite temperature}
%%%%%%%%%%%%%%%%%%%%%%%%%%%%%%%%%%%%%%%%%%%%%%%%%%%%%%%%%%%
%%%%%%%%%%%%%%%%%%%%%%%%%%%%%%%%%%%%%%%%%%%%%%%%%%%%%%%%%%%

%%%%%%%%%%%%%%%%%%%%%%%%%%%%%%%%%%%%%%%%%%%%%%%%%%%%%%%%%%%
\subsection{The model}
%%%%%%%%%%%%%%%%%%%%%%%%%%%%%%%%%%%%%%%%%%%%%%%%%%%%%%%%%%%

For a detailed introduction to field theories at finite temperature, we refer to
the books of \cite{kapusta} and of \cite{rothe}.
We first of all define the model under consideration.
In 3 dimensions at finite temperature $T$ (henceforth 3$d_T$ in short),
a field theory lives in a volume
of the toroidal shape
$T^{-1} \times {\bf R}^2$,
\be
   z = (z_0, z_1, z_2) \equiv (z_0,\vec{z}) \in {\bf R}/T \times {\bf R}^2,
\ee
cf. Fig.\, 2.
The extension is set to infinity in both spatial
directions, and the length of the 
torus component, the "temperature-direction",
is fixed by the inverse temperature.

%%%%%%%%%%%%%%%%%%%
% Fig. finite temperature torus
%%%%%%%%%%%%%%%%%%%
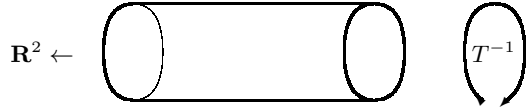
\begin{figure}
 
%\begin{center}
\setlength{\unitlength}{0.8cm}
\begin{picture}(10.0,2.0)
 
% diagram
 
\put(4.0,0.0){
\setlength{\unitlength}{0.8cm}
\begin{picture}(6.0,2.0)
 
% horizontal lines
{\linethickness{0.8pt}
\put(0.0,0.2){\line(1,0){4.0}}
\put(0.0,1.8){\line(1,0){4.0}}
}
% semicircles left
{\linethickness{0.8pt}
\qbezier(-0.5,1.0)(-0.5,0.2)(0.0,0.2)
\qbezier(-0.5,1.0)(-0.5,1.8)(0.0,1.8)
}
{\linethickness{0.1pt}
\qbezier(0.5,1.0)(0.5,0.2)(0.0,0.2) 
\qbezier(0.5,1.0)(0.5,1.8)(0.0,1.8) 
}
% semicircles right
{\linethickness{0.8pt}
\qbezier(3.5,1.0)(3.5,0.2)(4.0,0.2)
\qbezier(3.5,1.0)(3.5,1.8)(4.0,1.8)
}
{\linethickness{0.8pt}
\qbezier(4.5,1.0)(4.5,0.2)(4.0,0.2)
\qbezier(4.5,1.0)(4.5,1.8)(4.0,1.8)
}

% temperature circle
{\linethickness{0.8pt}
\qbezier(5.5,1.0)(5.5,0.4)(5.8,0.2)
\qbezier(5.5,1.0)(5.5,1.8)(6.0,1.8)
}
{\linethickness{0.8pt}
\qbezier(6.5,1.0)(6.5,0.4)(6.2,0.2)
\qbezier(6.5,1.0)(6.5,1.8)(6.0,1.8)
}
% arrows and T^{-1}
\put(6.2,0.18){\vector(-1,-1){0.1}}
\put(5.8,0.18){\vector(-1,1){0.1}}
\put(5.5,1.0){\makebox(1.0,0){$T^{-1}$}}
% R^2
\put(-2.0,1.0){\makebox(1.0,0){${\bf R}^2 \leftarrow$}}

\end{picture}
}
 
\end{picture}
 
%\end{center}
\caption{\label{finitet} The world of the $3d_T$ Gross-Neveu model.
The fermionic degrees of freedom are subject to anti-periodic
boundary conditions in the temperature direction.
}
 
\end{figure}
 
%%%%%%%%%%%%%%%%%%%
% end Fig. finite temperature torus
%%%%%%%%%%%%%%%%%%%

We consider a model of N Dirac spinor fields, that is of
N fermion species or flavours,
with values in a Grassmann (exterior) algebra,
\bea
   \psi(z) &=& (\psi_{\alpha i}(z))\vert_{\alpha=1,2; i=1,\dots ,N},
   \nonumber \\
   \overline\psi(z) &=& (\overline\psi_{\alpha i}(z))
    \vert_{\alpha=1,2; i=1,\dots ,N}.
\eea
Due to the trace operation of the partition function,
Grassmann-valued
fields are subject to anti-\-pe\-ri\-odic boundary conditions
in the temperature direction,
\be \label{antiper}
   \psi (z_0 \pm T^{-1}, \vec{z} ) = - \psi (z_0, \vec{z}),
\ee
and similarly for $\overline\psi$ (cf. e.g. \cite{luescher2}).
As a first consequence of the anti-periodicity,
the fermionic degrees of freedom always have an implicit
infrared cutoff involved, which is of the order of the
temperature $T$. This is easily seen by applying a
Fourier transform to momentum space,
\be \label{ftferm}
   \psi(z) = \; \sum_{k_0=\pi T (2n+1) , n \in {\bf Z}}
   e^{i k_0 z_0} \; \widetilde\psi_n(\vec{z}) .
\ee
From this we infer that $|k_0| \geq \pi T$,
which provides a lower bound on the total momentum.

The action is given by
\be \label{gnaction}
   S_{gn}(\psi, \overline\psi) = \int_z 
    ( \overline\psi(z) D \psi(z) )
   - \frac{\lambda^2}{N} \int_z \left(
      \overline\psi(z) \psi(z) \right)^2,
\ee
where
\bea
   \int_z & \equiv &
   \int_{{\bf R}^2} d^2\vec{z} 
   \int_{-\frac{1}{2}T^{-1}}^{\frac{1}{2} T^{-1}} dz_0 ,
   \nonumber \\
 \label{diff}
    D & = & \sum_{i=0}^2 \gamma_i \; \frac{\partial}{\partial {z_i}},
\eea
with $\gamma_0,\gamma_1,\gamma_2$ denoting the 3 basis elements
of a Dirac representation. We normalize them according to
\be
    \gamma_i \gamma_j + \gamma_j \gamma_i \; = \; 2 \delta_{ij}\; ; \;\;
   i,j = 0,1,2.
\ee
Here and in the following for spinor fields $\overline\psi(z)$,
$\eta(z)$ we use the notations
\be
    \overline\psi(z) \eta(z)
    \equiv \sum_{\alpha = 1,2} \; \sum_{i=1,\dots ,N}
      \overline\psi_{\alpha i}(z) \eta_{\alpha i}(z).
\ee
The partition function (generating functional of full correlation
functions) is then given by
\be \label{gnpartition}
   Z(\eta,\overline\eta) = \int \cD\overline\psi \cD\psi \;
   \exp{( - S_{gn}(\overline\psi,\psi)
   + \int_z ( \overline\psi(z) \eta(z) + \overline\eta(z) \psi(z)
   ) )},
\ee
where
\be \label{gnmeasure}
   \int \cD\overline\psi \cD\psi \;
    =  \prod_{z,\alpha,i} d\psi_{\alpha i}(z)
   d\overline\psi_{\alpha i}(z).
\ee
$\eta$ and $\overline\eta$ are external sources introduced
in order to derive correlation functions from
$Z(\eta,\overline\eta)$ by differentiation.

So far, the model is only formally defined by the action (\ref{gnaction})
and the partition function (\ref{gnpartition}).
In order to define it rigorously as a quantum field theory from the outset,
we have to introduce an ultraviolet cutoff $\Lambda$.
Among others, possible choices are a lattice, the Pauli-Villar
cutoff, dimensional regularization,
or just a simple momentum cutoff.
The model becomes well defined for $p\leq\Lambda$,
where $p$ represents momentum.
An essential point then is to prove the existence
of this model in the large cutoff limit $\Lambda\to\infty$,
and in this limit a sequence of axioms have to be satisfied
(\cite{osterwalder}, \cite{zinoviev}).
This is the issue of renormalization in the framework of
quantum field theory.
In many cases this is achieved by adjusting a
finite number of bare parameters of the action, fixing corresponding
renormalized coupling constants on a finite length scale.
Renormalizability implies that the influence of the cutoff
$\Lambda$ on the low-lying momentum scale $p$
is suppressed by some power of $p/\Lambda$.
In particular, all the correlation
functions of the model stay well-defined 
in the limit on finite scales.

One might argue that in the framework of statistical mechanics,
one is not necessarily interested in the removal of the ultraviolet cutoff.
For instance, the lattice spacing $a\sim 1/\Lambda$ is a finite
physical length in solid state physics
(except for the lattice spacing in the temperature direction,
which still is to be sent to zero).
However, concerning the critical regions of a model
where the correlation length diverges in units of the lattice
spacing, the motivation for renormalizability
is quite similar as in quantum field theory.
Instead of the cutoff being sent to infinity, 
the range of interest is
where the momentum becomes arbitrarily small.
Collective critical behaviour in this region to a large extent
becomes independent of the cutoff scale,
i.e. of the bare parameters.
This identifies a universality class,
that is universal critical behaviour.
This would not be true for non-renormalizable models.

In the following we regularize the Gross-Neveu model
by choosing a simple momentum cutoff $\Lambda$.
This is done mainly for computational simplicity.
Equally well we could have chosen a hybercubic lattice,
with some care concerning the
choice of lattice fermions.
 
There is some detailed knowledge of this model at zero
temperature $T=0$.
First of all, applying power counting of ultraviolet divergencies,
this 3-dimensional
model appears to be non-renormalizable in the loop expansion.
That is, as an expansion in powers of the coupling constant $\lambda$,
we need more and more couterterms with increasing order of $\lambda$
in order to render the ultraviolet limit finite.
Fortunately, for large number $N$ of fermionic flavours, the situation is
different.
Within the large $N$ expansion, which will be explained in the very detail
in the next section, the model becomes renormalizable, to every
finite order in $1/N$
(\cite{park1}).
Actually, there is an even stronger statement. Namely, if $N$ is
sufficiently large, the 3-dimensional Gross-Neveu model
can be shown to exist in the large cutoff limit
(and in the infinite volume),
that is, the correlation functions of the model stay finite.
The proof is done in the framework of constructive field
theory, using convergent expansions and the renormalization
group (\cite{daveiga1}, \cite{daveiga2}).
Remarkably enough, this is a construction
about a non-Gaussian fixed point,
leading to a non-trivial (interacting) quantum field theory.

Let us come back to the finite temperature.
As already said in the introduction, the model (\ref{gnaction})
serves as a good example for studying  a "chiral" phase
transition because it reveals essential similarities to massless QCD.
We seek for a symmetry that would be broken explicitly by
a fermionic mass term of the form
\be \label{fmass}
    m \overline\psi(z) \psi(z).
\ee
Actually, this will not be a true chiral symmetry. This would require
the existence of a $\gamma$-matrix, the $\gamma_5$, which anticommutes
with all generating elements ($\gamma_0, \gamma_1, \gamma_2$)
of a faithful representation
of the associated Clifford algebra
(cf. e.g. \cite{chevalley} or \cite{porteous}). It does not
exist in three dimensions. Instead,
the model is invariant under the global parity transformation
\bea \label{symm}
    \psi(z) & \to & \psi(-z) , \nonumber \\
    \overline\psi(z) & \to & - \overline\psi(-z).
\eea
Both the action (\ref{gnaction}) and the measure (\ref{gnmeasure})
stay invariant under this transformation.
The symmetry under (\ref{symm}) is spontaneously broken at $T=0$,
and it becomes restored at high temperature by a second order
phase transition, cf. below.
The parity condensate
\bea \label{porder}
   && \frac{1}{N} \; < \overline\psi(z) \psi(z) > 
    \nonumber \\
  && = \; \frac{1}{Z(0,0)}
  \int \cD\overline\psi \cD\psi 
   \; \frac{1}{N} \; \left( \overline\psi(z) \psi(z) \right)
    \exp{( - S_{gn}(\overline\psi, \psi ))}
\eea
serves as an order parameter of the transition.

We remark in passing that, besides others, the model has a
continuous global symmetry in flavour space, defined by
\be \label{onsymm}
    \psi_{\alpha i}(z)  \to  \sum_{j=1}^N U_{ij} \psi_{\alpha j}(z) 
    \; , \quad
    \overline\psi_{\alpha i} (z)  \to  \sum_{j=1}^N U^\dagger_{ji}
    \overline\psi_{\alpha j} (z),
\ee
with $U\in U(N)$.
This symmetry is neither explicitly broken by a mass term
(\ref{fmass}) nor does it get broken spontaneously.
In particular it yields fermion number conservation.

%%%%%%%%%%%%%%%%%%%%%%%%%%%%%%%%%%%%%%%%%%%%%%%%%%%%%%%%%%%
\subsection{Many flavours. The large $N$ expansion}
%%%%%%%%%%%%%%%%%%%%%%%%%%%%%%%%%%%%%%%%%%%%%%%%%%%%%%%%%%%
As mentioned in the previous section, much insight is obtained
by representing the Gross-Neveu model in a form that is
particularly well suited if the number of fermionic species is
sufficiently large.
The corresponding transformation is based on the resummation
of so-called planar graphs according to Fig.\, 3.

%%%%%%%%%%%%%%%%%%%
% Fig. resummation of planar graphs
%%%%%%%%%%%%%%%%%%%
\begin{figure}
 
%\begin{center}
\setlength{\unitlength}{0.8cm}
\begin{picture}(10.0,2.0)
 
% diagrams

\put(1.0,0.0){
\setlength{\unitlength}{0.8cm}
\begin{picture}(9.0,0.5)
 
% zero bubble
\put(0.0,1.0){\circle*{0.16}}
\put(0.0,1.0){\line(1,1){0.5}}
\put(0.0,1.0){\line(1,-1){0.5}}
\put(0.0,1.0){\line(-1,1){0.5}}
\put(0.0,1.0){\line(-1,-1){0.5}}

% +
\put(0.8,1.0){\makebox(0.1,0){$+$}}

% one bubble
\put(2.0,1.0){\circle{0.8}}
\put(1.6,1.0){\circle*{0.16}}
\put(2.4,1.0){\circle*{0.16}}
\put(2.4,1.0){\line(1,1){0.5}}
\put(2.4,1.0){\line(1,-1){0.5}}
\put(1.6,1.0){\line(-1,1){0.5}}
\put(1.6,1.0){\line(-1,-1){0.5}}

% +
\put(3.2,1.0){\makebox(0.1,0){$+$}}
                                                
% two bubbles
\put(4.4,1.0){\circle{0.8}}
\put(5.25,1.0){\circle{0.8}}
\put(4.0,1.0){\circle*{0.16}}
\put(4.82,1.0){\circle*{0.16}}
\put(5.65,1.0){\circle*{0.16}}
\put(5.65,1.0){\line(1,1){0.5}}
\put(5.65,1.0){\line(1,-1){0.5}}
\put(4.0,1.0){\line(-1,1){0.5}}
\put(4.0,1.0){\line(-1,-1){0.5}}

% + ... =
\put(6.5,1.0){\makebox(2.0,0){$+ \quad \cdots \quad =$}}

% sigma propagator
\put(9.3,1.0){\circle*{0.16}}
\put(10.9,1.0){\circle*{0.16}}
\put(10.9,1.0){\line(1,1){0.5}}
\put(10.9,1.0){\line(1,-1){0.5}}
\put(9.3,1.0){\line(-1,1){0.5}}
\put(9.3,1.0){\line(-1,-1){0.5}}
% semicircles up and down
{%\linethickness{0.8pt}
\qbezier(9.3,1.0)(9.3,1.2)(9.5,1.2)
\qbezier(9.5,1.2)(9.7,1.2)(9.7,1.0)
\qbezier(9.7,1.0)(9.7,0.8)(9.9,0.8)
\qbezier(9.9,0.8)(10.1,0.8)(10.1,1.0)

\qbezier(10.1,1.0)(10.1,1.2)(10.3,1.2)
\qbezier(10.3,1.2)(10.5,1.2)(10.5,1.0)
\qbezier(10.5,1.0)(10.5,0.8)(10.7,0.8)
\qbezier(10.7,0.8)(10.9,0.8)(10.9,1.0)

\put(10.1,1.3){\makebox(0.1,0){$\sigma$}}
}

\end{picture}
}
 
\end{picture}
 
%\end{center}
\caption{\label{planar} Resummation of planar graphs.
}
 
\end{figure}
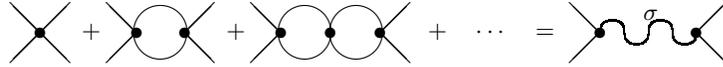
 
%%%%%%%%%%%%%%%%%%%
% end Fig. resummation of planar graphs
%%%%%%%%%%%%%%%%%%%

\be \nonumber
   \frac{\lambda^2}{N} \; \left(
   1 - c N \frac{\lambda^2}{N} +
   \left( c N \frac{\lambda^2}{N} \right)^2 + \cdots \right)
  =
  \frac{1}{N} \; \left( \frac{1}{\lambda^2} + c \right)^{-1}
\ee
Here $c$ counts the combinatorical factor and
the trace in spinor space over the single fermion bubble.
Such a chain then is represented as an interaction mediated
by an auxiliary scalar field $\sigma$.
 
We can do this resummation in closed form in the
following way.
First the local, quartic fermion interaction part of the action
can be written as the result of a gaussian integration, using
\bea \nonumber
   && \exp{ \left( \frac{\lambda^2}{N}
    \left( \overline\psi(z) \psi(z) \right)^2 \right) }
   \; = \;
   \left( \frac{N}{4\pi\lambda^2} \right)^{\frac{1}{2}} \cdot \\
   && \qquad \cdot \int_{-\infty}^\infty
   d\sigma(z) \;
   \exp{ \left( - \frac{1}{2} \left( \frac{N}{2\lambda^2} \right)
   \sigma(z)^2
   \pm \sigma(z) \; \overline\psi(z) \psi(z) \right)}.
\eea
The action then becomes of the Yukawa-type
and is bilinear in the fermionic fields.
In turn, the integration over the Grassmann algebra valued fields
can be done according to
\bea \nonumber
   && \int \cD\overline\psi \cD\psi \;
   \exp{ \left( - \int_z \int_{z^\prime} 
    \overline\psi(z) \; Q(z,z^\prime) \psi(z^\prime)
   + \int_z ( \overline\psi(z) \eta(z) + \overline\eta(z) \psi(z) )
   \right) } \\
  && \qquad = \det(-Q) \;
  \exp{\left( \int_z \int_{z^\prime} \overline\eta(z)
  Q^{-1}(z,z^\prime) \eta(z^\prime) \right)}.
\eea 
In our case, $Q$ is flavour diagonal, i.e. $Q = {\bf 1} \otimes K$,
with
\be \label{qform}
   K = D + \sigma \cdot 1
\ee
acting on the product of spinor and configuration space.
$D$ is defined by (\ref{diff}).
As the remnant of the anti-periodic boundary conditions 
(\ref{antiper}) imposed on the fermionic fields,
the momentum space representation of $D$ has
nonvanishing components only for the 
discrete thermal momenta
$k_0 = \pi T (2m+1)$, $m\in{\bf Z}$,
cf. (\ref{ftferm}).
In this way we obtain the following representation for the
partition function (\ref{gnpartition}) up to an unimportant
normalization constant.
\bea 
   Z(\eta,\overline\eta) &=& {\rm const} \; \int \cD\sigma \;
   \exp{ \left( - N \; S_{eff}(\sigma) \right) } \nonumber \\
  \label{effpartition}
   & \cdot &  \exp{ \left(  \int_z \int_{z^\prime} \overline\eta(z)
  ( 1 \otimes K^{-1} ) (z,z^\prime) \eta(z^\prime)
   \right)},
\eea
with
\be
   \int \cD\sigma \;
    =  \; \prod_{z} d\sigma(z).
\ee
The effective action $S_{eff}(\sigma)$ is given by
\be \label{effaction}
   S_{eff}(\sigma) \; = \;
   \int_z \left( \frac{\sigma(z)^2}{4\lambda^2}
   \; - \; \tr_s \left( \log{K} \right)(z,z)
   \right),
\ee
with the trace taken over the spinor indices.
A careful counting of the degrees of freedom
(e.g. by imposing a lattice cutoff) shows
that without loss of generality we may impose
periodic boundary conditions
on the auxiliary field $\sigma(z)$
in the temperature direction
(it is "bosonic").
This is the natural and common choice and implies
that constant field configurations of $\sigma$ are to be 
identified as zero momentum configurations
in the usual way.
In turn the minima of the effective action for large $N$
belong to the latter ones.

It should be emphasized that up to now no approximation
is involved. The partition function has only been rewritten
in a form that is most convenient if the number $N$ of fermions
becomes large. The only place where the number $N$ enters
in (\ref{effpartition}) is as the prefactor of the action.
The explicit form (\ref{effpartition}) reveals
the standard saddle point expansion
as the convenient method to study the Gross-Neveu model for large $N$.
The parity condensate (\ref{porder}) becomes the expectation value of
$-\tr_s K^{-1}(z,z)$,
and similar for the condensate correlations.
It is straightforward to show to all orders of $1/N$
that a vanishing condensate (\ref{porder}) is equivalent
to a vanishing expectation value of the auxiliary field $\sigma$.
The auxiliary field $\sigma(z)$ itself is parity odd.
The parity symmetry of the originial model becomes
invariance under
\be
   \sigma(z) \; \to \; -\sigma(-z).
\ee

\noindent
The large $N$ expansion proceeds along the following steps.
\begin{itemize}
\item Localization of the translation invariant minima of
the effective action $S_{eff}$. This yields the so-called
gap equation
\be \label{gapeqn}
   \frac{ \delta S_{eff}(\sigma) } { \delta \sigma(z) }
   \vert_{\sigma(z)\equiv\mu} \; = \; 0 \; !
\ee
\item Minimization requires positivity of
\be \label{gapstab}
   \frac{ \delta^2 S_{eff}}
   {\delta\sigma(z) \delta\sigma(z^\prime)}
   \vert_{\sigma(z) \equiv \sigma(z^\prime)\equiv\mu}.
\ee
\item Expansion about the solutions according to
\be \label{gapexpand}
   \sigma(z) \; = \; \mu \; + \; \frac{\phi(z)}{ N^{\frac{1}{2}} }.
\ee
\end{itemize}

Existence of stable solutions of the gap equation and 
their properties will be discussed in the next section.
For further reference we write down the model
as obtained by inserting the ansatz (\ref{gapexpand})
into the effective action (\ref{effaction})
and expand for large $N$.
Towards this end
we first introduce some notations used in the following.
Fermionic and bosonic thermal momenta are defined
as the sets
\bea
   \cF & = & \{ \frac{\pi}{\beta} ( 2 m + 1 ) \; \vert \; m \in {\bf Z} \},
   \nonumber \\
   \cB & = & \{ \frac{\pi}{\beta} 2 m \; \vert \; m \in {\bf Z} \}.
\eea
For the inverse temperature we write $\beta = T^{-1}$.
For $x, \beta\geq 0$ and $n=0, 1, 2, \dots$  let
\bea
   J_n(x,\beta) & = & \frac{1}{\beta} \sum_{q_0\in\cF}
   \int^\prime
   \frac{ d^2\vec{q} } { (2\pi)^2 } \;
   \frac{1}{(q^2 + x)^n}, \nonumber \\
  Q_n(x,\beta) & = & \frac{1}{\beta} \sum_{q_0\in\cF}
  \int^\prime
  \frac{ d^2\vec{q} } { (2\pi)^2 } \;
  \frac{\vec{q}^2}{(q^2 + x )^n}.
\eea
The prime at the integral sign denotes that the
sum and integral is confined to the region with
\be
   q^2 \equiv q_0^2 + \vec{q}^2 \; \leq \Lambda^2 .
\ee
For $\beta < \infty$ the $J_n(x,\beta)$ are always infrared
finite for every $x\geq 0$, whereas
\be
   J_n(x,\infty) = \int_{q^2\leq\Lambda^2}
   \frac{d^3q}{(2\pi)^3} \; (q^2 + x)^{-n}
\ee
is infrared singular as $x\to 0+$ for $n\geq 2$.
Furthermore, the ultraviolet singularities of the $J_n$ are
temperature independent,
which means that
\be
   \lim_{\Lambda\to\infty} \left(                                        
   J_n( x, \beta ) - J_n (x ,\infty ) \right) \; < \; \infty,            
\ee                                                                      
for each $x\geq 0$.
This is a property which is shared by a large class of 1-loop
Feynman integrals.
They can be rendered ultraviolet finite by imposing
the appropriate zero temperature normalization conditions.

Inserting (\ref{gapexpand})
into the effective action (\ref{effaction}),
we obtain the following large $N$ representation.
Let 
\be
    K_0 = D + \mu .
\ee
The partition function then becomes
\be                                                                                                                  
   Z(\eta,\overline\eta) = {\cN} \; \int \cD\phi \;
   \exp{ \left( - \; S(\phi)
   +  \int_z \int_{z^\prime} \overline\eta(z)
  ( 1 \otimes \widetilde{K}^{-1} ) (z,z^\prime) \eta(z^\prime)
   \right)},
\ee
with
\be \label{normal}
    \cN = {\rm const} \; \exp{\left( -N \int_z
    \left( 4\lambda^2 \mu^2 ( J_1(\mu^2,\beta) )^2
    - tr_s \log{(K_0)}(z,z)
    \right) \right) } ,
\ee
\be
   \widetilde{K}  = K_0 + \frac{\phi}{N^{\frac{1}{2}}},
\ee
and
\bea
   S(\phi) & = & \int_z \left( \frac{\phi(z)^2}{4\lambda^2}
    - N \; \tr_s \left( \log{( 1 + \frac{1}{N^{\frac{1}{2}}} \phi
    K_0^{-1} )}
    - \frac{1}{N^{\frac{1}{2}}} \phi K_0^{-1} \right)(z,z)
   \right)
   \nonumber \\
  & = & \sum_{n\geq 2} S_n(\phi).
\eea
Here,
\bea
   S_2(\phi) & = & \int_z \left( \frac{\phi(z)^2}{4\lambda^2}
   + \frac{1}{2} \; \tr_s \left( \phi K_0^{-1} \right)^2 (z,z) \right)
     , \nonumber \\
   S_n(\phi) & = & \frac{(-1)^n}{N^{n/2-1}} \; \frac{1}{n}\;
    \int_z \tr_s \left( \phi K_0^{-1} \right)^n(z,z) \; , \quad n\geq 3.
\eea
$S_n(\phi)$ is of the order of $O(N^{1-n/2})$.
For later convenience we write the quadratic part of the
action in its momentum space representation obtained
by Fourier transform.
With
\be
    \phi(x) =  \frac{1}{\beta} \sum_{k_0\in\cB}
    e^{i k\cdot x} \; \widetilde\phi(k)
\ee
we obtain after some algebra
\bea \nonumber
  S_2(\phi)  & = & \frac{1}{2} \;
     \frac{1}{\beta}\sum_{k_0\in\cB} \int^\prime
      \frac{d^2\vec{k}}{(2\pi)^2} \; \vert \widetilde\phi(k) \vert^2
     \biggl\{
     2 \left( \frac{1}{4\lambda^2} - J_1(\mu^2,\beta) \right) \\
   & + & \frac{1}{\beta}\sum_{q_0\in\cF}
     \int^\prime \frac{d^2\vec{q}}{(2\pi)^2}
     \biggl[
     \frac{k^2 + 4 \mu^2}{(q^2+\mu^2)((q+k)^2+\mu^2)}
    \\
   && \qquad\qquad\qquad\qquad 
    - \left(  \frac{1}{(q+k)^2+\mu^2} - \frac{1}{q^2+\mu^2} \right)
     \biggr]
   \; \biggl\}. \nonumber
\eea
The wave function renormalization constant $Z_R$
and the renormalized mass $m_R$
(the inverse correlation length) are defined by the small
momentum behaviour of the (connected) 2-point function
\be
    \widetilde{W}^{(2)}(k) \; = \;
    < \widetilde{\phi}(k) \widetilde{\phi}(-k) >
\ee
according to
\be
   \widetilde{W}^{(2)}(k_0=0,\vec{k}) \; = \;
   \frac{1}{Z_R} \left( m_R^2 + \vec{k}^2 + O(\vec{k}^4) \right)
  \;\; \mbox{as} \; \vec{k}\to 0.
\ee

%%%%%%%%%%%%%%%%%%%%%%%%%%%%%%%%%%%%%%%%%%%%%%%%%%%%%%%%%%%
\subsection{Phase structure in the $N=\infty$ limit}
%%%%%%%%%%%%%%%%%%%%%%%%%%%%%%%%%%%%%%%%%%%%%%%%%%%%%%%%%%%
We discuss the phase structure of the Gross-Neveu model
in the limit of infinite number of fermionic flavours first
(\cite{park2}).
The gap equation (\ref{gapeqn}) reads
\be \label{ngap}
   \mu \; \left( \frac{1}{4\lambda^2} - J_1(\mu^2,\beta) \right) \; = \; 0,
\ee
the solution of which yields the condensate $\mu(\beta)$.
We are mainly interested in the phase structure at finite temperature,
but it is instructive to include the case $\beta=\infty$ of
zero temperature.
$\mu = 0$ is always a solution of (\ref{ngap}).
For fixed $\beta$,
$J_1(\mu^2,\beta)$ is monotonically decreasing with
$\mu^2$.
Hence, for $1/(4\lambda^2) > J_1(0,\beta)$,
$\mu=0$ is the only solution, and it is 
straightforward to show that it is a minimum of the action.
On the other hand,
for $1/(4\lambda^2) < J_1(0,\beta)$
there are two non-trivial stable solution
$\sigma(z) = \pm \mu(\beta)$ of (\ref{gapeqn}) with
$\mu(\beta)>0$ satisfying (\ref{ngap}).

Now suppose that the bare coupling constant $\lambda$ is so large
that the parity-broken phase is realized at zero temperature.
We define a zero temperature mass scale $M$ by
\be
   J_1(M^2, \infty) \equiv \frac{1}{4\lambda^2} < J_1(0, \infty).
\ee
For fixed $\mu^2$, as a function of $\beta$,
$J_1(\mu^2,\beta)$ is monotonically increasing in $\beta$.
Decreasing $\beta$ down from $\beta = \infty$, there is a unique critical
value $\beta_c$ where the broken phase becomes unstable,
cf. Fig. 4.
This $\beta_c$ is given by
\be
   J_1(M^2,\infty ) = J_1(0, \beta_c)
  \quad\mbox{or}\quad \beta_c = \frac{2 \; \ln{2}}{M}.
\ee

%%%%%%%%%%%%%%%%%%%
% Fig. N=oo phase structure
%%%%%%%%%%%%%%%%%%%
\begin{figure}
 
%\begin{center}
\setlength{\unitlength}{0.8cm}
\begin{picture}(10.0,3.0)
 
% diagram
 
\put(2.0,0.5){
\setlength{\unitlength}{0.8cm}
\begin{picture}(9.0,0.5)
 
% horizontal axis
{%\linethickness{0.8pt}
\put(0.0,0.0){\line(1,0){6.0}}
\put(3.8,-0.5){\makebox(0.1,0){$T_c$}}
\put(6.0,-0.5){\makebox(0.1,0){$T=\beta^{-1}$}}

\put(4.0,1.5){\makebox(1.9,0){\sl high $T$ symmetric ($\mu=0$)}}

}
% vertical axis
{%\linethickness{0.8pt}
\put(0.0,0.0){\line(0,1){2.0}}
\put(-1.5,1.5){\makebox(1.0,0){$J_1(0,\infty)$}}
\put(-1.5,0.7){\makebox(1.0,0){$J_1(M^2,\infty)$}}
}
% curve
{%\linethickness{0.1pt}
\qbezier(0.0,1.5)(2.0,1.5)(5.0,0.2)
\put(0.5,1.7){\makebox(2.0,0){$J_1(0,\beta)$}}

}
% horizontal line
{%\linethickness{0.8pt}
\put(0.0,0.7){\line(1,0){0.6}}
\put(0.8,0.7){\line(1,0){0.6}}
\put(1.6,0.7){\line(1,0){0.6}}
\put(2.4,0.7){\line(1,0){0.6}}
\put(3.2,0.7){\line(1,0){0.6}}

\put(3.8,0.0){\line(0,1){0.3}}
\put(3.8,0.4){\line(0,1){0.3}}

\put(1.5,0.4){\makebox(1.7,0){\sl broken phase}}
}

\end{picture}
}
 
\end{picture}
 
%\end{center}
\caption{\label{ninfinity} Phase structure of the $3d_T$-dimensional
Gross-Neveu model at $N=\infty$. For given mass scale
$M$, the symmetric high temperature phase becomes unstable
at $T_c = M/(2\; \ln{2})$, and the model undergoes a second
order phase transition to the broken phase.
}
 
\end{figure}
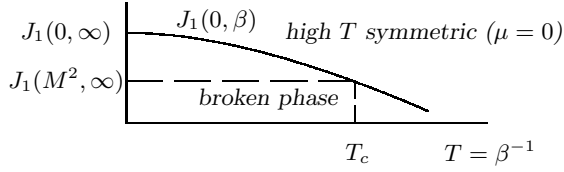
 
%%%%%%%%%%%%%%%%%%%
% end Fig. N=oo phase structure
%%%%%%%%%%%%%%%%%%%

In order to show that the symmetry gets actually broken
spontaneously for $\beta>\beta_c$,
before the infrared cutoff is removed (the volume is sent to
infinity)
we introduce an explicit parity breaking term (\ref{fmass})
into the original fermionic action (\ref{gnaction}).
Without loss we take $m>0$.
Then we remove the infrared cutoff and finally let
$m\to 0+$.
The only modification the fermionic mass term
implies for the derivation of the large $N$ representation
of the partition function as done in the last section is
the replacement of (\ref{qform}) by
\be
   K \; = \; D + ( m + \sigma ) \cdot 1 .
\ee
The gap equation (\ref{ngap}) is replaced by
\be \label{n2gap}
   ( \mu - m ) \; \frac{1}{4\lambda^2} - \mu J_1(\mu^2,\beta) \; = \; 0.
\ee
For small $m$ and $\beta>\beta_c$
there are two stable solutions of (\ref{n2gap}) given by
$\mu(m,\beta) = \pm \mu(\beta) + \delta$,
where $\mu(\beta)>0$ is the "unperturbed" solution of
(\ref{ngap}) and
\be
   \delta = \frac{1}{8\lambda^2 \mu(\beta)^2 J_2(\mu(\beta)^2,\beta)}
   \; m \; + \; o(m).
\ee
Inserting this into (\ref{normal}), we see that
the negative solution of the gap equation gets exponentially
suppressed in the thermodynamic limit for positive $m$.
We are left with a unique absolut minimum of the action,
breaking the parity symmetry.

The phase transition is of second order. The correlation length
$m_R^{-1}$ diverges at $\beta_c$.
We define critical exponents in the standard way, in particular
with $\chi_2$ as the 2-point susceptibility,
\bea
    \chi_2 \equiv \widetilde{W}^{(2)}(k=0)^{-1} & \simeq &
    (\beta_c - \beta)^{-\gamma} , \nonumber \\
   m_R & \simeq & (\beta_c - \beta)^\nu , 
   \qquad \mbox{as}\; \beta \to \beta_c- \\
   Z_R & \simeq & (\beta_c - \beta )^{\nu\eta},
   \nonumber
\eea
and similarly primed exponents $\nu^\prime, \gamma^\prime$
by approaching the phase transition from the broken phase.
To leading order in $N^{-1}$ the 2-point function is
given by
\bea
  \widetilde{W}^{(2)}(k_0=0,\vec{k}) & = & 
    2 \left( J_1(M^2,\infty) - J_1(0,\beta) \right)
    \nonumber \\
   &&\;\; + \; \vec{k}^2 \; 2 Q_3(0,\beta)
   + \; O(\vec{k}^4)  \qquad\mbox{(as $\vec{k}\to 0$)}  \\
   & = & \frac{\ln{2}}{\pi\beta\beta_c} \left( \beta_c - \beta \right)
    + \; \vec{k}^2 \frac{\beta_c}{16\pi} + O(\vec{k}^4)
       \nonumber
\eea
in the symmetric phase, and
\bea
  && \widetilde{W}^{(2)}(k_0=0,\vec{k}) = 
    4  \mu^2 J_2(\mu^2,\beta)
    \nonumber \\
   &&\;\; + \; \vec{k}^2 \; 2
     \left( J_2(\mu^2,\beta) - 2 \mu^2 J_3(\mu^2,\beta)
     + 4 \mu^2 Q_4(\mu^2,\beta) - Q_3(\mu^2,\beta) \right)
    \nonumber \\
   &&\;\; + \; O(\vec{k}^4)  \qquad\mbox{(as $\vec{k}\to 0$)}  \\
   && \;\; \simeq  \; \frac{2 \ln{2}}{\pi\beta_c^2} \left( \beta - \beta_c \right)
    + \; \vec{k}^2 \frac{\beta_c}{16\pi} + O(\vec{k}^4)
      \qquad\mbox{(as $\beta\to \beta_c-$)} \nonumber
\eea
in the broken phase.
We thus obtain
$\gamma = \gamma^\prime = 1$,
$\nu = \nu^\prime = 1/2$
and $\eta = 0$.
Other critical exponents are determined in
the straightforward way, with the result that
$\alpha = \alpha^\prime = 0$, $\delta = 3$
and $\beta = 1/2$.

%%%%%%%%%%%%%%%%%%%%%%%%%%%%%%%%%%%%%%%%%%%%%%%%%%%%%%%%%%%
\subsection{What happens for finite $N$?}
%%%%%%%%%%%%%%%%%%%%%%%%%%%%%%%%%%%%%%%%%%%%%%%%%%%%%%%%%%%
So far we know the phase structure of the
$3d_T$ Gross-Neveu model in the limit of infinite
number of flavours. As next, we are interested in the physics
of the model if there is a large but still finite number of
fermionic species.
The parity symmetry still remains broken at small
temperature and will be restored at sufficiently
high temperature.
Concerning the universality class of the proposed
second order transition two opposite
claims have been made.

On the one hand, one might follow the resoning proposed
by Pisarski and Wilczek.
The length of the temperature torus is given by the inverse temperature
$\beta$.
In the high temperature regime this length is rather small,
hence the geometry is essentially two-dimensional on
length scales large compared to $\beta$.
Furthermore, due to the anti-periodic boundary conditions to
be imposed on the fermionic degrees of freedom,
the latter are subject to an infrared cutoff, cf. (\ref{ftferm}).
Hence, in the far infrared, only the purely scalar modes survive.
It might be natural to assume that this provides a model that belongs
to the Ising universality class in two dimensions.
This would imply the following values for the critical exponents:
$\nu=1$, $\gamma=1.75$, $\eta =0.25$.
This is to be contrasted to the case of $N=\infty$.

On the contrary, the reasoning proposed by Kocic and Kogut
roughly goes as follows.
The model is subject to a $Z_2$ symmetry. 
This symmetry does not depend on the
number of fermionic flavours. For $N=\infty$ the model is gaussian
($\nu=1/2$, $\gamma=1$), hence it might be so for finite $N$.
A basic assumption made in this way of reasoning is that it is
the symmetry pattern which determines the universality
class of a phase transition. But we know that this can be
completely misleading. For instance, let us consider a classical spin 
model in 3 dimensions as described by an action of the type
$\phi^2 + \lambda \phi^4$. It has a global symmetry $\phi \to -\phi$
that does not depend on a particular value of the
quartic coupling $\lambda$. For $\lambda =0$ the model is purely 
Gaussian. But this is in sharp contrast to the critical behaviour 
at finite $\lambda$.
Similarly, for the Gross-Neveu model, the large $N$ behaviour
is not predicitve for finite $N$, 
at least not without closer inspection.

Which scenario is the right one?
The reduction argument as applied above is purely geometrical.
Does it generalize to field theory, and if so, how does the effective
2 dimensional model look like?
Renormalization group studies reveal an infinite number of
fixed points of those models, likely to be related to
various different universality classes.
It is not just the symmetry pattern alone that determines the
critical behaviour.
To decide the nature of the transition requires more
sophisticated methods known from quantum field theory.

Below we shall combine two techniques to resolve the puzzle
of the Gross-Neveu model.
The first one is dimensional reduction in the framework of
quantum field theories. This has become a well established
machinery in the recent past. It provides a computational
device to derive effective models for the quantitative
description of the infrared behaviour at high temperature.
The second method concerns convergent high temperature
series.
Under very general conditions, the high order behaviour
of those series allows for the study of the critical region,
including the measurement of critical exponents.

%%%%%%%%%%%%%%%%%%%%%%%%%%%%%%%%%%%%%%%%%%%%%%%%%%%%%%%%%%%
\section{\label{dr}Dimensional reduction in quantum field theory}
%%%%%%%%%%%%%%%%%%%%%%%%%%%%%%%%%%%%%%%%%%%%%%%%%%%%%%%%%%%
Field theories for realistic systems
and statistical models with a large number of
degrees of freedom evolve considerable complexity.
From the point of view of high energy physics,
good examples are quantum field theories such as
quantum chromodynamics, which is supposed to be the
microscopic theory of the strong interaction, and the
electroweak standard model.
The investigation of the large scale properties becomes very
demandingful.
In many instances
we are forced to use methods that imply appropriate simplifications.
On the one hand we will learn for the more complicated situation.
On the other hand it is desirable to have a systematic approach,
such as convergent or asymptotic expansions.
This allows for a quantitative
investigation and statements to be made for the full theory.

For field theories at finite temperature,
one of those systematic techniques is dimensional reduction.
Originally it served as a more or less qualitative reasoning
for the description of the infrared behaviour at high temperature.
It is easy to understand and works well for classical field theories.
The generalization to quantum field theories,
however, is neither simple nor obvious.
The first attempts were based on a naive generalization,
by just assuming that decoupling
properties of heavy masses that are well known
for zero temperature generalize to finite temperature.
It took some time until it was made definite that this
is not the case (\cite{landsman}).

A systematic way in which dimensional reduction
is realized in renormalizable quantized field theories
was given by the author (\cite{reisz1}).
Meanwhile it has become a very powerful technique.
In particular it has been used
to study the infrared properties of the QCD high temperature
phase, so as
to extract screening masses of the quark-gluon plasma
(\cite{petersson1}).
For a summary and further references,
cf. \cite{lacock}.
Later,
the success of this well developed technique and the experiences
obtained inspired its application
to the electroweak standard model of the early universe
(\cite{kajantie1}).
In the following we will shortly outline the ideas of
dimensional reduction and then will apply it to the Gross-Neveu model.

%%%%%%%%%%%%%%%%%%%%%%%%%%%%%%%%%%%%%%%%%%%%%%%%%%%%%%%%%%%
\subsection{A short survey of dimensional reduction}
%%%%%%%%%%%%%%%%%%%%%%%%%%%%%%%%%%%%%%%%%%%%%%%%%%%%%%%%%%%
We consider a local, renormalizable quantum field theory 
in thermal equilibrium  by the interaction with a heat bath,
in $D\geq 3$ dimensions.
In the functional integral formalism such a field theory
lives in a D-dimensional volume
with one dimension being compactified to a torus
of length given by the inverse temperature (Fig. 2).
There are boundary conditions imposed on the 
physical fields in this direction due to the
trace operation of the partition function.
Bosonic degrees of freedom are subject to periodic boundary conditions,
fermionic degrees of freedom to anti-periodic boundary conditions.

The idea behind dimensional reduction is based on the geometrical
picture that at high temperature $T$ the length of the temperature torus
given by $T^{-1}$ becomes small.
So as the temperature increases, the $D$-dimensional system becomes more
and more $D-1$-dimensional on spatial length scales $L$ large to
$T^{-1}$. In the following we discuss the implications for a
renormalizable quantum field theory.

Consider, for simplicity, a scalar model with field content $\phi(z)$,
described by the action
\be \label{draction}
   S(\phi) = \int_{0}^{T^{-1}} d z_0
    \int d^{D-1}\vec{z} \left( \frac{1}{2} \; \phi(z)
      \left( - \Delta_D\phi \right) (z)
   + V(\phi(z)) \right),
\ee
where $\Delta_D = \sum_{i=0}^{D-1} \partial^2/(\partial z_i^2)$
denotes the $D$-dimensional Laplacian with periodic boundary
conditions in $z_0$, and $V$ the self-interaction of the field.
Periodicity implies that the corresponding momentum components
in this direction are discrete.
We apply a so-called Matsubara decomposition, which is nothing
but a Fourier transform with respect to this
momentum component,
\bea \label{matsubara}
   \phi(z) & = & T^{1/2} \; \sum_{k_0=2 n \pi T, n \in {\bf Z}}
   e^{i k_0 z_0} \; \widetilde\phi_n(\vec{z}) \nonumber \\
   & = & T^{1/2} \; \phi_{st}(\vec{z}) \; + \;
   \phi_{ns}(z),
\eea
where the static field $\phi_{st}$ denotes the $n=0$
component and the non-static field $\phi_{ns}$ sums
all components with $n\not=0$.
The kinetic part of the action becomes
\bea
    \int d z_0 && \int d^{D-1}\vec{z} \; \frac{1}{2} \;
    \phi(z) \left( - \Delta_D\phi \right) (z) \nonumber \\
   &&= \int d^{D-1}\vec{z} \; \frac{1}{2} \;
     \widetilde\phi_0(\vec{z})
     \left( - \Delta_{D-1}\widetilde\phi_0 \right) (\vec{z}) \\
   &&\; +  \sum_{n \not= 0} \int d^{D-1}\vec{z} \; \frac{1}{2} \;
        \widetilde\phi_n(\vec{z}) 
     \left(
     \left( - \Delta_{D-1}\widetilde\phi_n \right) (\vec{z})
     + (2\pi n T)^2 \widetilde\phi_n(\vec{z})
    \right) . \nonumber
\eea
Apparently the field components with $n \not= 0$ acquire a
thermal mass proportional to $T$. This provides an infrared cutoff
to these non-static modes mediated by the interaction with
the heat bath.
The only modes which do not acquire a $T$-dependent mass term are
the $\widetilde\phi_0$. They are called static
modes because they are precisely those field components that do
not fluctuate in $T$-direction.
They are the only degrees of freedom that survive dynamically on scales
$L\geq T^{-1}$.
All the other modes decouple. We call this scenario 
{\sl complete dimensional reduction}.

So far, the consideration is done on the level of the action, that is,
for the classical field theory. How does this picture change under
quantization? Now we have to investigate in full 
the partition function given by
\be
   Z \; = \; \int \cD\phi \; \exp{(-S(\phi))} \quad ; \quad
   \cD\phi = \prod_z d\phi(z),
\ee
with the action $S(\phi)$ as in (\ref{draction}).
From the Matsubara-decomposition (\ref{matsubara}) it appears that
the finite temperature field theory 
if the $\phi(z)$ in $D$ dimensions can be viewed as
a $D-1$-dimensional field theory
of the $\widetilde\phi_n(\vec{z})$ at zero temperature.
There are multiple fields $\widetilde\phi_n$ now almost all of which
become massive. The only effect of the temperature is to renormalize the
interaction.

One might guess that on the infrared scale the heavy modes decouple
as in the classical case, the only degrees of freedom left being
the static fields $\widetilde\phi_0(\vec{z})\equiv\phi_{st}(\vec{z})$.
There are well known statements in quantum field theory 
on large mass behaviour. 
The Ambj\o rn-Appelquist-Carazonne theorem
predicts the decoupling of massive fields from the low-energy scale
of the theory under quite general conditions.
(It was proposed originally by \cite{appelquist}.
The proof of this theorem uses the BPHZ renormalization scheme and
is due to \cite{ambjorn}.)
In many cases it applies, such as for the simple scalar field theory above.
The statement is very specific for the renormalization scheme used.
All the coupling constants of the model have to be defined on the low
momentum scale.
In other instances the theorem does not apply.
This concerns models with an anomaly
or those subject to a non-linear symmetry,
in which case the correlation functions are constrained by Ward identities.

The obstruction that prevents the application of this theorem in our case
is peculiar. The number of heavy modes is infinite, whereas the above
theorem does not make any prediction uniform on the number of modes.
Anyway it would be a surprise if this theorem applies to field theories
at finite temperature. For there is a strong interplay between
renormalization and decoupling properties.
Its validity would allow to circumvent the issue of
ultraviolet renormalization of the theory,
which is an intrinsic D-dimensional problem.

Whether or not the quantum fluctuations destroy complete dimensional
reduction must be decided by other means.
The answer was given by \cite{landsman}
by means of a careful perturbative renormalization group analysis.
Apart from some exceptional cases (such as QED),
the answer is negative.
To understand this fact, we have to account for renormalization
theory.
Let $\widetilde\Gamma(\vec{p},T)$ denote a generic
connected or
1-particle irreducible correlation function,
$\vec{p}$ a generic spatial momentum.
$\widetilde\Gamma$ achieves contributions both from the static and
the non-static modes.
The analogy of the classical decoupling property of the latter
reads
\be \label{complete}
   \widetilde\Gamma(p,T) \; = \; \widetilde\Gamma_{st}(p,T)
   \left( 1 + O(\frac{p}{T},\frac{\mu}{T}) \right),
\ee
where $\widetilde\Gamma_{st}$ denotes the purely static
contribution to $\widetilde\Gamma$.
$\mu$ denotes some finite mass parameter.
For renormalizable field theories we can always achieve this
in the first instance,
by imposing normalization conditions on some low momentum
scale. These conditions define the physical, renormalized
coupling constants of the theory.
The bare parameters of the model, that is the coupling constants
of the action, are functions of the physical coupling constants
and the ultraviolet cutoff. They are defined in such a way that
the correlation functions $\widetilde\Gamma$
stay finite in the large cutoff limit and become functions of the
renormalized coupling constants and the temperature only.
This is the issue of renormalization.

The point to be made now is that the bare parameters
do not depend on the temperature $T$. The latter only enters
through the torus of length $T^{-1}$.
In order to obtain (\ref{complete}), the subtractions must be
temperature dependent, and so are the renormalized parameters.
In particular, the renormalized mass $\mu$ that enters (\ref{complete})
becomes of the order of $T$ in the generic case.
Hence the correction term becomes of the order of 1.
This shows that complete dimensional reduction breaks down.

In accordance to the decomposition (\ref{matsubara})
of the fields into static and non-static components,
let us write the partition function as
\be
   Z \; = \; \int \cD\phi_{st} \; \exp{(-S_{eff}(\phi_{st}) )}
\ee
with
\be
   \exp{(-S_{eff}(\phi_{st}) )} \; = \;
   \int \cD\phi_{ns} \;
   \exp{(-S(T^{1/2}\phi_{st}+\phi_{ns}) )}
\ee
obtained by integrating out the non-static modes.
Formally we can write this as
\be
   S_{eff}(\phi_{st}) \; = \;
   S_0(\phi_{st}) + (\delta_{ns}S)(\phi_{st}),
\ee
where $S_0$ is the classically reduced action of the temporal zero
modes $\phi_{st}$.
This action is completely super-renormalizable.
For the example above, it is given by
\be
    S_0(\phi_{st}) \; = \; \int d^{D-1}\vec{z} \;
    \left( \frac{1}{2} \;
    \phi_{st}(\vec{z}) \left( - \Delta_{D-1} \phi_{st} \right) (\vec{z})
    + V(\phi_{st}(\vec{z})) \right).
\ee
The quantum fluctuations of the non-static modes are completely
described by the correction term $\delta_{ns}S$.
This induced interaction is non-local.
We know from the discussion above
that it cannot be neglected even at very high temperature.
Nevertheless one would still think of an appropriate high temperature expansion.
Not any expansion will work because we have to avoid the same obstruction
leading to (\ref{complete}).

The solution is to generate the high temperature
reduction as the low momentum expansion (\cite{reisz1}).
$\delta_{ns}S$ is made up of purely non-static modes.
Thus it has an intrinsic mass gap. Its momentum representation
is analytic at zero momentum $\vec{p}$.
We are thus allowed to expand about $\vec{p}=0$.
In a renormalizable field theory, the high temperature behaviour
of the renormalized $n$-point function
$\widetilde\Gamma_{ns}^{(n)}$
is proven to be
%**% renormalizability implies p/Lambda << p/T
\be
   \widetilde\Gamma_{ns}^{(n)}(\vec{p},T) \; = \;
   T^{\rho_n} P_{\rho_n}(\frac{\vec{p}}{T})
   \; + \; o(1)
   \;\mbox{as}\; T\to\infty .
\ee
$P_m$ denotes a (multidimensional) polynomial of degree $m$ for $m\geq 0$,
$P_m=0$ for $m<0$,
and
\be \label{irpc}
   \rho_n \; = \; (D-1) - n \frac{D-3}{2}.
\ee
The degree $\rho_n$ depends only on $n$ (and $D$), but
is independent of the order to which
$\widetilde\Gamma_{ns}^{(n)}$ has been computed in a perturbative
(e.g. weak coupling) expansion.

Let us write $T_{\vec{p}}^\delta f(\vec{p})$ for the (multidimensional)
Taylor expansion about $\vec{p}=0$ to the order $\delta$
of the function $f(\vec{p})$.
To keep the finite $T$ contributions on the infrared scale
amounts to replace for those $n$ with $\rho_n\geq 0$
\be \label{expansion}
   \widetilde\Gamma_{ns}^{(n)}(\vec{p},T) \; \to \;
   T_{\vec p}^{\rho_n - q_n}
   \widetilde\Gamma_{ns}^{(n)}(\vec{p},T)
\ee
and put all others to 0.
Here, $q_n$ denotes some integer number with 
$0\leq q_n\leq \rho_n$.
We notice that the expansion does not generate ratios of the
form $\mu/T$. The expansion is applicable even for $T$-dependent
normalization conditions where the renormalized masses $\mu$ 
increase with $T$.
 
Only non-static correlations $\widetilde\Gamma_{ns}^{(n)}$
with $\rho_n\geq 0$ contribute to $\delta_{ns}S$ at high
temperature. The contribution is polynomial in momentum
space, that is, $\widetilde\Gamma_{ns}^{(n)}$
generates a local $n$-point interaction of $\delta_{ns}S$.
Hence we end up with a locally interacting model of
the fields $\phi_{st}(\vec{z})$.

If we put $q_n=0$ above, i.e. keep the complete high $T$ 
part of the non-static amplitudes,
the effective model becomes a (power counting-) 
renormalizable field theory in $D-1$ dimensions.

We may be interested to keep only the leading
term for every correlation function.
In this case we put $q_n = \rho_n$ in (\ref{expansion}) and
obtain
\be \label{partexp}
   \widetilde\Gamma_{ns}^{(n)}(\vec{p},T) \; \to \;
   T_{\vec p}^0
   \widetilde\Gamma_{ns}^{(n)}(\vec{p},T)
  \equiv \widetilde\Gamma_{ns}^{(n)}({\bf 0},T)
\ee
for all amplitudes $\widetilde\Gamma_n$
with $\rho_n\geq 0$.
This restriction is meaningful at sufficiently
large spatial separations compared to the inverse 
temperature $T^{-1}$, for instance for the investigation of
universal critical behaviour at a second order phase transition.
Corrections are suppressed by the order of $\vec{p}/T$.

We should emphasize at this point that (\ref{partexp}) is not equivalent
to keeping only those effective interactions
that stay super-renormalizable
(except for the reduction from $D=3$ to $D-1=2$
dimensions, cf. below).
In general, in addition this would require to omit the contributions
from the non-static correlation functions
$\widetilde\Gamma_{ns}^{(n)}$
with $\rho_n=0$. 
For example, for the reduction from 4 to 3 dimensions this would imply to
ignore interactions of the form $\phi^6$.
This is justified only if the temperature is
sufficiently large. Without further justification this omission
is a potential danger
for the investigation of the
finite temperature phase transition. 

If only the effective interactions mediated by the non-static
modes are kept which do preserve super-renormalizability
of the (D-1)-dimensional effective model, the latter is made ultraviolet
finite by the $D$-dimensional counterterm of the original
theory, of course projected in the same way as in (\ref{partexp}).
In the more general case, supplementary normalization conditions
have to be imposed. They are obtained by matching conditions
to the D-dimensional theory.

Let us summarize the state of the art for a local renormalizable
quantum field theory at finite temperature.
Its infrared behaviour on spatial length scales  $L\geq T^{-1}$
is quantitatively described by a local, generically renormalizable or even
super-renormalizable field theory 
of the purely static modes $\phi_{st}(\vec{z})$.
This is a field theory in one less dimension and at zero temperature.
The non-static degrees of freedom $\phi_{ns}$, which account for the
oscillations along the temperature torus, 
have an intrinsic mass gap of the order of $T$ and hence
do not survive as dynamical
degrees of freedom in this region.
However, they induce local interactions of the static fields
that inevitably have to be taken into account in order to describe 
the infrared properties in a correct way.
This concerns both the properties of the high temperature phase
as well as finite temperature phase transitions.
From the point of view of statistical mechanics,
precisely these interactions are responsible for
the phenomenon that
quantum critical behaviour can be different from the critical behaviour
of the corresponding classical model.
Which interactions actually have to be kept depends on the situation
at hand, as explained above.

A remarkable property of dimensional reduction is the fact that fermions
do not survive the reduction process as dynamical degrees of freedom.
As discussed in the last chapter, the anti-periodic boundary conditions
imposed on the fermion fields in the $T$-direction provide a mass
gap, identifying fermions as purely non-static modes.
The "only" effect they have on the infrared scale is mediated by
renormalization of the interaction of the otherwise bosonic fields.

The actual computation of the effective action of the static
fields $\phi_{st}$ can be done by various means.
For four-dimensional gauge theories, QCD and the electroweak standard model
weak coupling renormalized perturbation theory has been
used extensively.
For the Gross-Neveu model
below we will carry out the dimensional reduction
by means of the large $N$ expansion.

%%%%%%%%%%%%%%%%%%%%%%%%%%%%%%%%%%%%%%%%%%%%%%%%%%%%%%%%%%%
\subsection{Application to the three-dimensional Gross-Neveu model
at finite temperature}
%%%%%%%%%%%%%%%%%%%%%%%%%%%%%%%%%%%%%%%%%%%%%%%%%%%%%%%%%%%
We are back now to the Gross-Neveu model. 
This model is renormalizable for a large number $N$ of flavours.
Hence, dimensional reduction as presented in the last section applies.
We have $D=3$, so by (\ref{irpc}) we get $\rho_n$=2 for all $n$.
All non-vanishing correlation functions
$\widetilde\Gamma_{ns}^{(n)}(\vec{p},T)$ grow at most as $T^2$ at high
temperature.

We are interested in the parity restoring phase transition
and apply the high temperature expansion 
in the form (\ref{partexp}) to all non-static effective vertices.
That is, we keep the leading high temperature behaviour
of $\widetilde\Gamma_{ns}^{(n)}(\vec{p},T)$ for all $n$.
It is convenient to rescale the twodimensional fields by a factor
of $(16\pi)^{1/2}/\beta$.
Then the effective action is given by
\bea \label{gneffaction}
   S_{eff}(\phi) =
    \int \frac{d^2\vec{z}}{(2\pi)^2} \;
      & \biggl( &
      \frac{1}{2} m_0^2 \phi(\vec{z})^2
      + \frac{1}{2} \sum_{i=1}^2 
        \left( \frac{\partial}{\partial z_i} \phi(\vec{z})\right)^2
      \nonumber \\
      & + &  \sum_{m\geq 2}
      (-1)^m \frac{c_m}{N^{m-1}\beta^2}
      \phi(\vec{z})^{2m}
     \biggr) .
\eea
Here and in the following we write $\phi$
instead of $\phi_{st}$ for the static fields.
Only even powers of $\phi$ occur in the effective action.
The bare mass squared $m_0^2$ and the constants $c_m$ for $m\geq 2$
are given by
\bea \label{gneffcouplings}
    m_0^2 & = & \frac{16 \ln{2}}{\beta^2\beta_c} \; ( \beta_c - \beta )
    \; + \; \frac{\overline{m}^2}{\beta^2 N} \; + \; O(\frac{1}{N^2})
  \nonumber \\
  c_m & = & \frac{1}{m} \; \frac{J_m(0,\beta)}{\beta^{2m-3}} \;
   (16\pi)^m \; ( 1 + O(\frac{1}{N}) ).
\eea
The $c_m$ do not depend on the number of flavours.
Also, they are temperature independent non-negative numbers
because of $0 \leq J_m(0,\beta) = {\rm const}\; \beta^{2m-3}$,
with $m$-dependent constant. In particular, we have
\bea
    c_2 & = & 8 \; \pi , \nonumber \\
    c_3 & = & \frac{32}{9} \pi^2 , \\
    c_4 & = & \frac{128}{75} \pi^3 . \nonumber
\eea
If $N=\infty$, only the quadratic part of the action survives
and the model is purely Gaussian.
On the other hand, for large but finite $N$, which we are interested
in, we obtain an interacting model.
The interactions are alternating in sign.
For every $m\geq 2$,
the $2m$-point coupling constant is of the order
of $N^{1-m/2}$. For large $N$
interactions $\phi(\vec{z})^{2m}$ with large $m$ are highly suppressed.

The ${\bf Z}_2$-symmetric model described by the action (\ref{gneffaction}) is
super-\-re\-nor\-maliz\-able. The $1/N$ expansion is an expansion
in the number of loops.
The only ultraviolet divergence occurs for the 2-point function
to the one loop order, which is the order $N^{-1}$.
There is only one corresponding, logarithmically divergent,
mass counterterm included. This is the second term for $m_0^2$
of (\ref{gneffcouplings}).
Its precise value does not play a role for the universality class
of the phase transition,
as will be seen in the next chapter.
All other correlation functions stay ultraviolet finite and are not
subject to subtractions.

%%%%%%%%%%%%%%%%%%%%%%%%%%%%%%%%%%%%%%%%%%%%%%%%%%%%%%%%%%%
\section{\label{finiten}Phase transition for finite number of flavours $N$}
%%%%%%%%%%%%%%%%%%%%%%%%%%%%%%%%%%%%%%%%%%%%%%%%%%%%%%%%%%%
We are prepared now to investigate the parity restoring phase transition
of the Gross-Neveu model. This will be done by
determining the phase structure of the associated 2-dimensional
effective model described by the action (\ref{gneffaction}).

Two-dimensional field theories typically evolve a rich and complicated
phase structure.
For instance, renormalization group studies of hierarchical
models reveal that
${\bf Z}_2$ symmetric scalar models have complicated fixed point
manifolds in field space (\cite{pordt1}).
We have to expect more than just one (Ising) universality class for
2d scalar models, depending to a large extent on their interactions
and the coupling constant strengths.

The interactions and the values of the coupling constants of
the effective model (\ref{gneffaction}) have been computed by
dimensional reduction.
They are given functions of the number $N$ of flavours and
of the temperature $T=\beta^{-1}$.
The model is superrenormalizable, and there is only one
logarithmically cutoff dependent mass counterterm.
This is the only term that depends on the particular cutoff chosen.
%**% All the other coupling constants are "universal".
For the following non-perturbative investigation we put the model
on the lattice.

First of all, we truncate the action beyond the interaction $\phi^8$.
Terms of the order of $N^{-4}$ are omitted.
This is legitimate for finite, but large $N$, which we are interested in.
On the lattice, the action becomes
\bea
   a^2 \sum_{\vec{x}\in\Lambda_2} & \biggl\{ &
   \left( \frac{m_0^2}{2} \phi(\vec{x})^2
   + \frac{1}{2} \sum_{\mu=1,2} ( \widehat\partial_\mu\phi)(\vec{x}) ^2 \right)
   \nonumber \\
   & + & T^2 \left( \frac{c_2}{N} \phi(\vec{x})^4
    - \frac{c_3}{N^2} \phi(\vec{x})^6
    + \frac{c_4}{N^3} \phi(\vec{x})^8 \right)
   \biggr\}.
\eea
Here $\Lambda_2$ denotes the two-dimensional hypercubic lattice
with lattice spacing $a$, and $\widehat\partial$ is the lattice
derivative
\be
   (\widehat\partial_\mu\phi)(\vec{x}) \; = \;
   \frac{1}{a} \left( \phi(x + a \widehat\mu) - \phi(\vec{x}) \right),
\ee
$\widehat\mu$ as unit vector in the positive $\mu$-direction.

The choice of the lattice cutoff is consistent.
For, the lattice model is (super-)renormalizable.
According to well known power counting theorems on the lattice
(\cite{reisz0}),
the continuum limit where the lattice spacing is sent to zero exists
to all orders of the $1/N$ expansion.
That is, all correlation functions of the basic field $\phi$
stay ultraviolet finite. Furthermore, in this large cutoff limit they agree
with those of the model
with e.g. the simple momentum cutoff, which we have chosen so far.

For further use we write the action  in a form that is well known for
lattice spin models. Furthermore, the lattice spacing is put to $1$
in the following.
We rescale the fields by introducing the so-called hopping parameter
$\kappa$ and define further ultralocal coupling constants
$\lambda$, $\sigma$ and $\omega$
by
\bea
   \phi(\vec{x}) & = & (2\kappa)^{1/2} \; \phi_0(\vec{x}) , \nonumber \\
   \kappa ( 4 + m_0^2 ) & = & 1 - 2 \lambda + 3 \sigma - 4 \omega ,
   \nonumber \\
   4 c_2 \kappa^2 \frac{T^2}{N} & = & \lambda - 3 \sigma + 6 \omega , \\
    - 8 c_3 \kappa^3 \frac{T^2}{N^2} & = & \sigma - 4 \omega , \nonumber \\
    16 c_4 \kappa^4 \frac{T^2}{N^3} & = & \omega . \nonumber
\eea
Then, up to an irrelevant constant normalization
factor, we obtain the partition function in the form
\be \label{spinz}
   Z \; = \; \int \prod_{\vec{x}\in\Lambda_2}
   \; \exp{( - S_0(\phi_0))}
\ee
with the action
\bea \label{spinaction}
   S_0(\phi_0)) & = & \sum_{\vec{x}\in\Lambda_2}
   \left( S^{(0)}(\phi_0(\vec{x})) - (2\kappa) \sum_{\mu=1,2}
   \phi_0(\vec{x}) \phi_0(\vec{x}+\widehat\mu) 
    \right), \nonumber \\ && \nonumber \\
   S^{(0)}(\phi_0(\vec{x})) & = & \phi_0(\vec{x})^2
   + \lambda ( \phi_0(\vec{x})^2 - 1 )^2  \\
   && + \sigma ( \phi_0(\vec{x})^2 - 1 )^3
   + \omega ( \phi_0(\vec{x})^2 - 1 )^3 . \nonumber
\eea
The coupling constants $\lambda$, $\sigma$ and $\omega$
are of the order of $N^{-1}$, $N^{-2}$
and $N^{-3}$, respectively, and are given by
\bea \label{spincouplings}
   \lambda & = & 36 \; \left( 1 - \frac{4}{\cR} + \frac{288}{25\cR^2}
    \right) \bar{l}, \nonumber \\
   \sigma & = & -\frac{48}{\cR} \; \left( 1 - \frac{144}{25\cR}
    \right) \bar{l} , \\
   \omega & = & \frac{1728}{25} \; \frac{1}{\cR^2} \;
   \bar{l}, \nonumber
\eea
with
\be
   \bar{l} = \frac{8\pi \kappa^2 T^2}{9N}
   \qquad\mbox{and}\qquad
   \cR = \frac{3N}{2\pi\kappa}.
\ee
The particular values where simultaneoulsy
$\bar{l}=0$ and $\cR=\infty$ correspond
to the case of infinite number of flavours, $N=\infty$. The model then is a
free Gaussian model. For all other values
we introduce the ratios
\bea \label{alphabeta}
    \widehat\alpha & = & 
     \frac{\lambda}{\omega} \; = \; \frac{25}{48} \cR^2 \;
    \left( 1 - \frac{4}{\cR} + \frac{288}{25\cR^2}
    \right) , \nonumber \\
   \widehat\beta & = & \frac{\sigma}{\omega} \; = \; 
    -\frac{25}{36} \cR \; \left( 1 - \frac{144}{25\cR}
    \right) .
\eea
As we will see, the phase structure can be parametrized as
function of the ratios $\widehat\alpha$ and $\widehat\beta$.

The finite temperature phase transition and its properties will be
determined non-perturbatively by means of convergent
high temperature series.
We will apply the linked cluster expansion (LCE).
It provides hopping parameter series for the free energy
density and the connected correlation function of the model
about completely disordered lattice systems.
It should be emphasized that LCE generates convergent
power series,
convergence being uniform in volume.
This implies that correlations can be measured 
with arbitrary precision
within the domain of convergence,
the precision depending on the order of computation.
Even more, under quite general conditions,
the phase transition is identified with the radius of
convergence.
The critical behaviour of the model is encoded in the high
order coefficients of the series.
This concerns both the location of the transition
as well as the critical exponents and amplitudes.

The hopping parameter series have been generated recently
for two-, four, and six-point functions up
to the 20th order, for arbitrary coupling constants
(\cite{reisz2}).
The series are available for general O(N) scalar field models
on hypercubic lattices in $D\geq 2$ dimensions.
The lattices are unbounded in all directions or they have
one compact dimension, representing the temperature
torus.
Also, the expansion has been generalized to finite lattices in order to
allow for the determination of the order of a transition
by finite size scaling analysis
(\cite{hildegard2}).
This LCE technique allows for a rapid generation of the series
representation for all coupling constants.
We will outline the idea behind this in the next section.

%%%%%%%%%%%%%%%%%%%%%%%%%%%%%%%%%%%%%%%%%%%%%%%%%%%%%%%%%%%
\subsection{Convergent series expansion - LCE}
%%%%%%%%%%%%%%%%%%%%%%%%%%%%%%%%%%%%%%%%%%%%%%%%%%%%%%%%%%%
Hopping parameter expansions 
of classical lattice spin systems
have a long tradition in statistical physics.
Good references are \cite{wortis} and \cite{itzykson}.
In contrast to generic perturbation theory about eigenstates of
harmonic oscillators or free fields, which are normally at best
asymptotically convergent, such expansions provide convergent 
series about disordered lattice systems.

Linked cluster expansion (LCE) is a hopping parameter expansion
applied to the free energy density, connected correlation functions and
susceptibilities.
Providing absolutely convergent power series,
under quite general conditions
they allow for the determination of the critical properties of a
statistical model.
This includes critical couplings as well as critical exponents and amplitudes.
These numbers are encoded in the high order behaviour of
the series coefficients. Their determination
requires a computation to high orders.

The realization of such expansions by convenient algorithms
and with the aid of computers have been
poineered by \cite{luescher0}. 
They were the first to perform such
expansions to the 14th order of the hoppping parameter.
Recent progress and generalizations have been made
in various directions.
They have been generalized to field theories at finite temperature
respectively quantum spin systems, where one dimension is
compact with torus length given by the inverse temperature
(Reisz (1995)).
To the finite volume, allowing for finite size scaling analysis
in order to determine the nature of a phase transition
(\cite{hildegard2}).
In the course of this construction it was possible to
prolong the length of the series to the order 20.
Other progresses concern classical fixed-length spin systems
(\cite{butera1})
or higher moments of two-dimensional scalar models (\cite{campostrini1}).

Below
we only give a short outline of the ideas behind the expansion technique.
Details will be given only to the extent they are needed for the application
in the next section.

Let $\Lambda_D$ denote a
$D$-dimensional hypercubic lattice.
For simplicity we work with units where the lattice spacing becomes 1.
We consider a class of O(N) symmetric scalar fields
living on the sites of the lattice. The partition function is given by
\be \label{lcepart}
   Z(H,v) = \int \prod_{x\in\Lambda_D} d^N\Phi(x) \;
   \exp{(-S(\Phi,v)+\sum_{x\in\Lambda} H(x)\cdot\Phi(x) )}.
\ee
Here $\Phi = (\Phi_1,\dots , \Phi_N)$
denotes a real, $N$-component scalar field, $H$ are
external (classical) fields, and
\[
    H(x)\cdot\Phi(x) \; = \; \sum_{a=1}^N H_a(x)\Phi_a(x).
\]
The action is given by
\be \label{lceaction}
   S(\Phi,v) =  \sum_{x\in\Lambda_D} S^0(\Phi(x))
   - \frac{1}{2}\;
   \sum_{x\not=y\in\Lambda_D}
   \sum_{a,b=1}^N \Phi_a(x)v_{ab}(x,y)\Phi_b(y).
\ee
The "ultra-local" (0-dimensional) part $S^0$ of the action
describes the single spin measure. We assume that it is
$O(N)$ invariant and that it ensures stability of the partition function
\eqn{lcepart}.
For instance,
\be \label{lceultra}
   S^0(\Phi) \; = \; \Phi^2 + \lambda (\Phi^2-1)^2
   + \sigma (\Phi^2-1)^3 + \omega (\Phi^2-1)^4,
\ee
with $\omega>0$. Dependence on the
bare ultra-local
coupling constants $\lambda$, $\sigma$ and $\omega$ is suppressed in the
following. Fields at different lattice sites are coupled by the 
hopping term $v(x,y)$. For the case of nearest neighbour interactions,
\be \label{lcehopping}
   v_{ab}(x,y) \; = \; 
    \delta_{a,b} \; \left\{
   \begin{array}{r@{\qquad ,\quad} l }
    \; 2\kappa\; & {\rm x,y \; nearest\; neighbour,} \\
    0 & {\rm otherwise.}
   \end{array} \right.
\ee
$\delta$ is the Kronecker symbol.

The generating functional of connected correlation functions
(free energy) is given by
\bea \label{lcefree}
  W(H,v) & = & \ln{ Z(H,v) }, \\
   && \nonumber \\
  \label{lcecorr}
  W_{a_1\ldots a_{2n}}^{(2n)} (x_1,\ldots, x_{2n}) & = &
   <\Phi_{a_1}(x_1) \cdots \Phi_{a_{2n}}(x_{2n}) >^c
   \nonumber \\
   & = & \left.
   \frac{\partial^{2n}}{\partial H_{a_1}(x_1) \cdots
   \partial H_{a_{2n}}(x_{2n})}
    W(H,v) \right\vert_{H=0}.
\eea
In particular, the connected 2-point susceptibility $\chi_2$ and the moment $\mu_2$
are defined according to
\bea \label{lcesuscepts}
  \delta_{a,b}\; \chi_2 & = &
   \sum_x < \Phi_a(x)\Phi_b(0) >^c , \nonumber \\
 && \nonumber \\
  \delta_{a,b}\; \mu_2 & = &
   \sum_x \left( \sum_{i=0}^{D-1} x_i^2 \right)
    < \Phi_a(x)\Phi_b(0) >^c .
\eea
The Fourier transform 
$\widetilde{W}(p)$ of the 2-point function is defined by
\be
    W_{a_1 a_2}^{(2)} (x_1, x_2) 
   = \delta_{a_1, a_2} \int_{-\infty}^\infty
     \frac{d^D p}{(2\pi)^D} \;
     e^{i p\cdot (x_1-x_2)} \;
    \widetilde{W}(p).
\ee
The standard definitions of the renormalized mass $m_R$
(as inverse correlation length) and the
wave function renormalization constant $Z_R$
are by the small momentum behaviour of
$\widetilde{W}$,
\be 
  \widetilde{W}(p)^{-1} \; = \; \frac{1}{Z_R} \,
   ( m_R^2 + p^2 + O(p^4) ) \quad
   {\rm as} \;p \to 0 .
\ee
From (\ref{lcesuscepts}) we obtain the relations
\be \label{lcerenorm}
  m_R^2 = 2 D \frac{\chi_2}{\mu_2} \; , \;
  Z_R = 2 D \frac{\chi_2^2}{\mu_2} .
\ee
The critical exponents $\gamma,\nu,\eta$ are defined by the
leading singular behaviour at the critical point $\kappa_c$,
\bea
  \ln{\chi_2} & \simeq & -\gamma \ln{(\kappa_c-\kappa)}, \nonumber \\
  \label{lceexpo}
  \ln{m_R^2} & \simeq & 2\nu \ln{(\kappa_c-\kappa)}\; , \quad
 \mbox{as}\; \kappa\nearrow\kappa_c , \\
  \ln{Z_R} & \simeq & \nu\eta \ln{(\kappa_c-\kappa)}, \nonumber
\eea
such that
$\nu\eta=2\nu-\gamma$.

Let us assume for the moment that
the interaction part (\ref{lcehopping}) of the action
between fields at different lattice sites
is switched off, i.e.~$v=0$. Then the action decomposes
into a sum of the single spin actions over the lattice sites,
$S(\Phi, v=0)=\sum_x S^0(\Phi(x))$, and the partition
function factorizes according to
\be
    Z(H, v=0)= \prod_{x\in\Lambda_D} Z^0(H(x)).
\ee
It follows that
\be
  W(H, v=0)= \sum_{x\in\Lambda_D} W^0(H(x)) ,
\ee
with $W^0 = \ln{Z^0}$.
In particular, the connected correlation functions
(\ref{lcecorr}) vanish except for the case that
all lattice sites agree,
i.e.~for $x_1=x_2=\cdots=x_{2n}$. In that case,
\be \label{lcecorrultra}
  W_{a_1\ldots a_{2n}}^{(2n)} (x_1,\ldots, x_{2n}) \; = \; 
    \frac{\stackrel{\circ}{v}_{2n}^c}{(2n-1)!!} C_{2n}(a_1,\ldots , a_{2n}).
\ee
The $C_{2n}$ are the totally symmetric
O(N) invariant tensors
with $C_{2n}(1,\dots, 1)$ $=$
$(2n-1)!! \equiv 1\cdot 3 \cdots \dots
\cdot (2n-1)$.
We have introduced the vertex couplings as defined by
\be \label{lcevc}
  \stackrel{\circ}{v}_{2n}^c = \left.
   \frac{\partial^{2n}}{\partial H_1^{2n}}\; W^0(H)
   \right\vert_{H=0}.
\ee
The vertex couplings
$\stackrel{\circ}{v}_{2n}^c$ 
depend only on the ultralocal part $S^0$ of the action,
that is on the single spin measure.
They are obtained from the relation $W^0(H) = \ln{Z^0(H)}$ or
\be \label{lcevcc}
    W^0(H) =  \sum_{n\geq 1} \frac{1}{(2n)!}
    \stackrel{\circ}{v}_{2n}^c (H^2)^n
   =  \ln{(1+\sum_{n\geq 1}  \frac{1}{(2n)!}
    \stackrel{\circ}{v}_{2n} (H^2)^n )},
\ee
where
\be \label{lcev2n}
     \stackrel{\circ}{v}_{2n} = \frac{ \int d^N\Phi \;
       \Phi_1^{2n} \exp{(-S^0(\Phi))} }
       { \int d^N\Phi
        \exp{(-S^0(\Phi))} },
\ee
by comparing coefficients of the Taylor expansions about $H=0$.
The resulting relations of the $\{\stackrel{\circ}{v}_{2n}^c\}$
and the $\{\stackrel{\circ}{v}_{2n}\}$ is one-to-one.
To some extent, the $\stackrel{\circ}{v}_{2n}$
encode universality classes,
as we will see below. 

The linked cluster expansion for $W$ and connected
correlations $W^{(2n)}$ is the
Taylor expansion with respect to the hopping couplings
$v(x,y)$ about this completely decoupled case,
\be \label{lceexpansion}
  W(H,v) = \left. \left( \exp{\sum_{x,y} \sum_{a,b} v_{ab}(x,y)
   \frac{\partial}{\partial \widehat v_{ab}(x,y)}} \right) W(H,\widehat v)
   \right\vert_{\widehat v =0}.
\ee
The corresponding expansions of correlation functions are
obtained from \eqn{lcefree} and \eqn{lcecorr}.
Multiple derivatives of $W$ with respect to
$v(x,y)$ are computed by the generating equation
\be \label{lcegenerate}
  \frac{\partial}{\partial v_{ab}(x,y)} W(H) = \frac{1}{2} \left(
   \frac{\partial^2 W}{\partial H_a(x) \partial H_b(y)} +
   \frac{\partial W}{\partial H_a(x)} \frac{\partial W}{\partial H_b(y)}
   \right) .
\ee

The expansion (\ref{lceexpansion}) is convergent
under the condition that the
interaction $v(x,y)$ is sufficiently local and weak
(cf. \cite{pordt2} and \cite{pordt3}).
Convergence is uniform in the volume.
In all that follows we restrict attention to the case of (\ref{lcehopping}),
that is to euclidean symmetric nearest neighbour interaction of strength
$\kappa$.
By inspection, for the susceptibilities such as (\ref{lcesuscepts})
the coefficients of the power series
of $\kappa$ are of equal sign. This remarkable property
implies that the singularity closest to the origin lies on the positive
real $\kappa$-axis. This singularity then is 
the phase boundary or the critical point.
In turn it is possible to extract quantitative information on
behaviour even close to criticality from the high order
coefficients of the series.

We know that e.g. the 2-point susceptibility (\ref{lcesuscepts})
is represented by the convergent series expansion
\be
   \chi_2 \; = \; \sum_{L\geq 0} a_{L,2} \; (2\kappa)^L
\ee
with real coefficents $a_{L,2}$.
From the high order behaviour of the coefficients of the series
we extract the critical point $\kappa_c$ and
the singular critical behaviour of $\chi_2$ close to $\kappa_c$. As
\be \label{lcesing}
   \chi_2 \simeq \left( 1 - \frac{\kappa}{\kappa_c} \right)^{-\gamma},
\ee
we have
\be \label{lceratios}
   \frac{a_{L,2}}{a_{L-1,2}} \; = \;
  \frac{1}{2\kappa_c} \left( 1 + \frac{\gamma-1}{L} + o(L^{-1}) \right).
\ee
Analogous relations hold for the other correlations, as for the correlation
length $m_R^{-1}$ and for $Z_R^{-1}$.
Similarly, critical amplitudes are extracted.
There are more sophisticated and also more involved methods than
(\ref{lceratios}) to obtain the critical numbers to high precision.
Our application below is to distinguish Ising and mean field
behaviour in two dimensions, which have rather different critical exponents.
For this the ratio criterion (\ref{lceratios}) 
turns out to be sufficiently precise.

The explicit computation of the series becomes rapidly complex with increasing
order. It is mainly a combinatorical problem and is conveniently handled
by graph theoretical terchniques. Correlation functions are represented
as a sum over equivalence classes of connected graphs,
each class being endowed with the appropriate weight it contributes.
For instance, 
the coefficients $a_{L,2}$ are obtained as the sum over
the set $\cG_{L,2}$ of connected graphs with $2$ external and
$L$ internal lines,
\be \label{lcecoeff}
   a_{L,2} \; = \; \sum_{\Gamma\in\cG_{L,2}} w(\Gamma),
\ee
with $w(\Gamma)$ the weight of the graph $\Gamma$.
Let $\cV_\Gamma$ denote the set of vertices of $\Gamma$,
and for every $v\in\cV_\Gamma$ let $l(v)$ be the sum of
internal and external lines of $\Gamma$ attached to the vertex $v$.
The weight $w(\Gamma)$ then is given by
\be \label{lceweight}
   w(\Gamma) \; = \; \cR(\Gamma) \cdot
   \prod_{v\in\cV_\Gamma} 
   \frac{\stackrel{\circ}{v}^c_{l(v)}(S^0)}{(l(v)-1)!!} .
\ee
The first factor $\cR(\Gamma)$ is a rational number. It is the product of the
(inverse) topological symmetry number of $\Gamma$,
the internal O(N) symmetry factor, and the
lattice embedding number of the graph $\Gamma$.
This is the number of ways $\Gamma$
can be put onto the lattice under the constraints imposed by the graph
topology, appropriately supplemented with possible geometrical weights,
as e.g. for $\mu_2$ of (\ref{lcesuscepts}).
Two vertices have to be put onto nearest neighbour lattice sites
if they have at least one internal line in common. Beyond this there are no
further exclusion constraints. In particular, a lattice site may be convered
by more than just one vertex (free embedding).

The second factor of (\ref{lceweight}) is the product over all vertices $v$
of $\Gamma$ of the vertex couplings $\stackrel{\circ}{v}^c_{l(v)}(S^0)$.
This is the only place where the ultralocal coupling constants of $S^0$
enter. Generally this factor is a real number, 
it becomes rational only in exceptional cases.
An example of a weight $w(\Gamma)$ is given in Fig.~5.

%%%%%%%%%%%%%%%%%%%
% figure: weight of a graph example
%%%%%%%%%%%%%%%%%%%

\begin{figure}

%\begin{center}
\setlength{\unitlength}{0.8cm}
\begin{picture}(10.0,2.0)

% diagrams

\put(0.0,0.0){
\setlength{\unitlength}{0.8cm}
\begin{picture}(9.0,0.5)

% one bubble
\put(2.0,1.0){\circle{1.0}}
\put(1.5,1.0){\circle*{0.16}}
\put(2.5,1.0){\circle*{0.16}}
\put(1.5,1.0){\line(1,0){1.0}}
% another bubble
\put(3.1,1.0){\circle{1.0}}
\put(2.55,1.0){\circle*{0.16}}
\put(3.55,1.0){\circle*{0.16}}
\put(2.55,1.0){\line(1,0){1.0}}
% external lines
\put(1.55,1.0){\line(-1,0){0.5}}
\put(3.55,1.0){\line(1,0){0.5}}

% weight
\put(5.5,1.0){\makebox(7.0,0.0){
$\left(\frac{1}{3! 3!}\right)\left( 2D\right)^2
\left( 9 (N+2)(N+4) \right)
\; \left( \frac{1}{3} \stackrel{\circ}{v}_4^c \right)^2
\left( \frac{1}{15} \stackrel{\circ}{v}_6^c \right) $}}
                                                                                                          
\end{picture}
}

\end{picture}

%\end{center}
\caption{\label{weight} A graph contributing to the connected 2-point function
of the $D$-dimensional O(N) model,
and its weight.
}

\end{figure}
%%%%%%%%%%%%%%%%%%%
% end figure: weight of a graph example
%%%%%%%%%%%%%%%%%%%

For given dimension, lattice topology and internal symmetry,
$\cR(\Gamma)$ is a fixed rational number.
Its computation is expensive, but needs to be done only once.
Changing the parameters of the action (that is, $\kappa$ and $S^0$)
does not require to compute $\cR(\Gamma)$  a new.
$S^0$ only enters the vertex couplings $\stackrel{\circ}{v}^c_{2n}$.
Once we have computed all the $\cR(\Gamma)$, the computation of the
LCE series for a new $S^0$ requires only a minimal amount of computing time.

The number of graphs $\Gamma$ and in turn of the $\cR(\Gamma)$ is large.
To the 20th order there are more than $10^8$ numbers $\cR(\Gamma)$.
Actually, it is not necessary to keep the $\cR(\Gamma)$ for all graphs
$\Gamma$ of interest.
The idea to circumvent this is to introduce the so-called
vertex structures.
Roughly speaking, a vertex structure accounts for the numbers
of vertices with particular numbers of lines attached.
More precisely,
a vertex structure $w$ is a sequence of non-negative integers,
\be \label{lcevs}
    w \; = \; ( w_1, w_2, \dots ) \quad , \quad
    w_i \in \{ 0,1,2,\dots \},
\ee
with only finitely many $w_i\not= 0$.
We write $\cW$ for the set of all vertex structures.
For a graph $\Gamma$ the vertex structure $w=w(\Gamma)$
$\in\cW$ associated to $\Gamma$ is defined by
(\ref{lcevs}) with
$w_i$ equal to the numbers of vertices of $\Gamma$
that have precisely $i$ lines attached.
With the definition
\be
    \cR(w) \; = \; \sum_{\Gamma\in\cG_{L,2} \atop
    w(\Gamma) = w }
   \cR(\Gamma)
\ee
we obtain instead of (\ref{lcecoeff}) and (\ref{lceweight})
\be
   a_{L,2} \; = \; \sum_{w\in W} \cR(w) \;
   \prod_{i\geq 1} \left( \frac{\stackrel{\circ}{v}^c_{i}}{(i-1)!!} \right)^{w_i} .
\ee
The sum is finite because most of the $\cR(w)$ vanish.
Actually there are only
a couple of hundreds of non-trivial $\cR(w)$'s,
to be compared to $10^8$ $\cR(\Gamma)$'s.

In order to make high orders in the expansion feasible it is necessary
to introduce more restricted
subclasses of graphs such as 1-line and
1-vertex irreducible graphs and renormalized moments. The correlations
are then represented in terms of the latter.
The major problems to be solved are algorithmic in nature. 
Examples are:
\begin{itemize}
\item unique algebraic representation of graphs
\item construction and counting of graph
\item computing weights, in particular lattice embedding numbers.
\end{itemize}
High orders are feasible by largely exploring simplifications.
On the hypercubic lattice, only graphs matter that have an even number of
lines in each loop. For O(N) symmetric models, vertices with an odd
number of lines attached vanish.
The most important series are available to the
20th order yet.

%%%%%%%%%%%%%%%%%%%%%%%%%%%%%%%%%%%%%%%%%%%%%%%%%%%%%%%%%%%
\subsection{The nature of the parity transition}
%%%%%%%%%%%%%%%%%%%%%%%%%%%%%%%%%%%%%%%%%%%%%%%%%%%%%%%%%%%
We determine the phase transition of the model
(\ref{spinz}), (\ref{spinaction}).
Before we put the coupling constants to their prescibed values
given by (\ref{spincouplings}),
it is instructive to discuss the strong coupling limit first.

As we have mentioned in the last section,
universality classes can be read off to some extent
from the $n$-dependence of the vertex couplings
$\stackrel{\circ}{v}_{2n}$ as defined in (\ref{lcev2n}),
without having to go through the complete analysis of the LCE series.
For ${\bf Z}_2$-symmetric models,
the two cases we are interested in are the following.
Let $z$ be any positive real number. Then
\be \label{vgauss}
          \stackrel{\circ}{v}_{2k} \; = \; \frac{(2k-1)!!}{2^k} \; z^k
\ee
for all $k=0,1,2,\dots$ implies that the model belongs to the universality
class of the Gaussian model.
The critical hopping parameter $\kappa_c$ is given by
$\kappa_c z = 1/(2D)$ in $D$ dimensions,
and the critical exponents are
$\gamma=1$, $\nu=1/2$ and $\eta=0$.
On the other hand, let the vertex couplings be given by
\be \label{vising}
          \stackrel{\circ}{v}_{2k} \; = \; \frac{(2k-1)!!}{2^k} \; z^k
     \; \frac{\Gamma(\frac{1}{2})}{\Gamma(\frac{1}{2}+k)}.
\ee
Then the universality class is that of the Ising model.
In two dimensions, $\kappa_c z =(1/4)\ln{(2^{1/2}+1)}$,
$\gamma=1.75$, $\nu=1$ and $\eta=0.25$.

The strong coupling limit of (\ref{spinz}), (\ref{spinaction})
is defined by $\omega\to\infty$ with the ratios
$\widehat\alpha = \lambda/\omega$ and $\widehat\beta=\sigma/\omega$ hold fixed.
The behaviour of the vertices  $\stackrel{\circ}{v}_{2k}$
is obtained by the saddle point expansion.
As function of $\widehat\alpha$ and $\widehat\beta$ the model evolves
a rather complicated phase structure.
We don't need to discuss it in detail, but the following
regions are of particular interest to us.

There are regions where the behaviour is Gaussian,
and regions where it is Ising like.
For $\widehat\alpha/\widehat\beta^2 > 1/4$ we obtain (\ref{vising}) with
$z=1$. In this case, $\kappa_c = (1/4)\ln{(2^{1/2}+1)}$
independent of $\widehat\alpha$ and $\widehat\beta$.
For $0<\widehat\alpha/\widehat\beta^2 < 1/4$
and $\widehat\beta<0$ we again get
(\ref{vising}), with
$z= 1-(3\widehat\beta/8)\left( 1 + \left( 
1 - \frac{32}{9}\frac{\widehat\alpha}{\widehat\beta^2} \right)^{1/2} \right)>0$.
In both cases the transition is Ising like.
On the other hand, 
in the region $0<\widehat\alpha/\widehat\beta^2 < 1/4$ and $\widehat\beta>0$,
$\widehat\alpha< \min{( \frac{3}{2}\widehat\beta-2, \widehat\beta-1 )}$
implies (\ref{vgauss}), with $z\sim\omega^{-1}$.
The model is Gaussian behaved or completely decouples over
the lattice because of $z\to 0$.
In Fig.~5 we show a qualitative plot of the 
universality domains.

%%%%%%%%%%%%%%%%%%%
% Fig. phase plot
%%%%%%%%%%%%%%%%%%%
\begin{figure}[htb]

\begin{center}
\setlength{\unitlength}{0.8cm}
\begin{picture}(10.0,3.0)
 
% diagram
 
\put(1.0,0.0){
\setlength{\unitlength}{0.8cm}
\begin{picture}(8.0,3.0)

% coordinate axis
{\linethickness{0.6pt}
\put(0.0,0.0){\line(1,0){8.0}}
\put(3.0,0.0){\line(0,1){3.0}}
\put(4.0,-0.5){\makebox(1.0,0){$\widehat\beta$}}
\put(2.2,2.0){\makebox(1.0,0){$\widehat\alpha$}}
}
% alpha = 1/4 * beta^2
\qbezier(3.0,0.0)(4.5,0.0)(8.0,3.0)
% upper bound on Gaussian
\put(5.0,0.75){\line(2,1){2.5}}
\put(4.5,0.0){\line(2,3){0.5}}
% phase transition regions  
\put(4.0,1.5){\makebox(1.0,0){I}}
\put(4.5,1.5){\circle{0.5}}
\put(1.5,1.5){\makebox(1.0,0){I}}
\put(2.0,1.5){\circle{0.5}}
\put(6.0,0.5){\makebox(1.0,0){II}}
\put(6.5,0.5){\circle{0.5}}

\end{picture}
}

\end{picture}
\end{center}
 
\caption{\label{critstruct}
Part of the phase structure of the effective model
in the strong coupling limit. 
In the regions I the critical exponents are those of the 2-dimensional
Ising model. Region II is the Gaussian domain or decoupled across
the lattice.}

\end{figure}
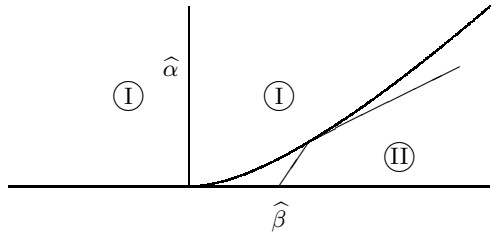
 
%%%%%%%%%%%%%%%%%%%
% end Fig. phase plot
%%%%%%%%%%%%%%%%%%%

In our case the ratios $\widehat\alpha$ and $\widehat\beta$ 
are largely fixed. They are given by (\ref{alphabeta}), and
\be
    \frac{\widehat\alpha}{\widehat\beta^2} \; = \; \frac{27}{25} \;
    \frac{1-\frac{4}{\cR} + \frac{288}{25\cR^2}}
   { \left( 1 - \frac{144}{25\cR} \right)^2 } .
\ee
Because $\cR$ is proportional to $N$, for sufficiently large
$N$ we have both $\widehat\beta<0$ and $\widehat\alpha/\widehat\beta^2>1/4$.
This inevitably selects the Ising universality class.

Next we consider the case of finite coupling constants
$\lambda$, $\sigma$ and $\omega$. We invoke
the LCE to construct the large order series of the susceptibilities
$\chi_2$ and $\mu_2$ as defined in (\ref{lcesuscepts})
for the various coupling constants.
From the series coefficients then we extract the critical coupling
$\kappa_c$ and the critical exponents $\nu$ and $\gamma$,
according to (\ref{lcesing}) and (\ref{lceratios}) and in analogy for
the correlation length $m_R^{-1}$.

%**% kappa_c -kappa \sim T - T_c 

We proceed in the following way. We choose $\cR = O(N)$ sufficiently
large. Then $\widehat\alpha$ and $\widehat\beta$ are fixed. 
For every $\cR$ we vary the 8-point coupling $\omega$ from
$\omega = \infty$ down to small values, the other couplings
running according to $\lambda = \widehat\alpha \omega$ and
$\sigma = \widehat\beta\omega$.
From the LCE series of $\chi_2$ we obtain the critical
hopping parameter $\kappa_c(\lambda,\sigma,\omega)$.
This then determines the precise value of $N$ 
corresponding to the transition point by
\be
    N \; = \; \frac{ 2 \pi \kappa_c}{3} \; \cR.
\ee
Then the critical exponents $\gamma$ and $\nu$ are measured.
As a result, their values are independent of $\omega$, and they are
equal to the Ising model ones.
The exception is a small region about 
the Gaussian model with $\lambda=\sigma=\omega=0$,
where there is a smooth crossover to the Gaussian numbers.
This is to be ascribed as a truncation effect of the finite order series.
The above process is repeated for various values of $\cR$
so that $N$ varies over a large region.

In Table \ref{critexp} and Fig. 7 we have collected some numbers
on the critical exponents $\gamma$ and $\nu$ obtained in this way.
For all $N$ the critical exponents are those of the 2-dimensional
Ising model. 
For small $N$ this might be not trustworthy because we have obtained
the effective model within the large $N$ expansion.
On the other hand, for large $N$ this is predictive. 

%%%%%%%%%%%%%%%%%%
% table : critical exponents
%%%%%%%%%%%%%%%%%%

\begin{table}[htb]

\vspace{0.5cm}

\begin{center}
\begin{tabular}{|c|c|c|}
\hline \hline
$\quad$ N $\quad$ & $\quad \gamma(N) \quad $ &
$ \quad 2 \nu(N) \quad $   \\
[0.5ex] \hline
 46.1  & 1.751(2)  & 2.001(2)     \\
 24.3  & 1.752(3)  & 2.001(4)      \\
 4.68  & 1.749(2)  & 1.999(2)       \\ [0.5ex] \hline
 1.41  & 1.750(2)  & 2.001(2)    \\\hline
\end{tabular}
 
\end{center}
\caption{\label{critexp}
The critical exponents $\gamma$ and $\nu$ of the three-dimensional
finite temperature Gross-Neveu model for various numbers $N$ of flavours.
They agree to those of the two-dimensional Ising model.
}

\end{table}

As a result, we conclude that the 3-dimensional Gross-Neveu model,
with large but finite number $N$of fermionic flavours,
has a finite temperature phase transition that belongs to the
universality class of the 2-dimensional Ising model.

%%%%%%%%%%%%%%%%%%%%%%%%%%%%%%%%%%%%%%%%%%%%%%%%%%%%%%%%%%%
\section{Summary}
%%%%%%%%%%%%%%%%%%%%%%%%%%%%%%%%%%%%%%%%%%%%%%%%%%%%%%%%%%%
We have investigated  the finite temperature phase structure of the
three-dimensional Gross-Neveu model.
It is a parity symmetric, four-fermion interacting model with $N$ fermion
species, and we kept $N$ large but finite.

From the point of view of high energy physics,
the interest in this model is based on the observation that it
reveals remarkable similarities to 
the chiral phase transition of QCD with two massless flavours.
Being a three-dimensional model, chirality is replaced by
parity, but both symmetries would be broken explicitly by a fermionic
mass term.

Beyond its similarity to QCD, the advantage of this model is that
to a large extent we have a closed analytic control of it.
This includes the finite temperature transition.

At zero temperature, parity symmetry is spontaneously broken.
It becomes restored at some critical temperature $T_c$
by a second order transition.
There are at least two candidates of universality classes this
transition might belong to. These are the two-dimensional
Ising model and mean field behaviour.
Both scenarios could be true,
following different conventional wisdom arguments.
It became clear in this lecture that the definite answer can be given only by
explicit computations.

Our result is that for large but finite number $N$ of fermions,
the transition belongs to the universality class of the two-dimensional
Ising model, with
critical exponents $\gamma =1.75$, $\nu=1$ and $\eta=0.25$.
This result is obtained by the combination of three computational techniques:
Large $N$ expansion, dimensional reduction in quantum field theory
and high order linked cluster expansions.

The large $N$ representation of the Gross-Neveu model amounts to a
resummation of planar graphs, that is, of chains of fermion bubbles,
cf. Fig.~3.
In the path integral approach this can be achieved in closed form
by representing the four-fermi interaction as the result of a
Yukawa-type interaction with an auxiliary scalar field.
Whereas the weak (quartic) coupling expansion is not renormalizable,
the model becomes (strictly) renormalizable for large $N$,
in particular in the expansion in powers of $N^{-1}$.

Dimensional reduction is to be considered as a special high temperature
expansion that is applied in the course of integrating out the degrees
of freedom that fluctuate along the temperature torus.
It is applied to a quantum field theory
(rather than a classical field theory)
to describe its infrared properties.
Applied to the Gross-Neveu model, the latter is mapped
onto a two-dimensional effective field theory.
Due to the anti-periodicity of the fermionic fields,
the model is purely bosonic.
It is local and super-renormalizable.
The interactions of this model are computed within the large $N$ expansion.
Both because the transition at the critical temperature $T_c$
is of second order and $N$ is supposed to be
sufficiently large, dimensional reduction will work down to $T_c$.

Finally, the relevant phase structure of the dimensionally reduced
effective model is determined. This is done by applying
the linked cluster expansion.
This technique allows us to compute convergent hopping parameter series
for the free energy and connected correlation functions
to very high orders.
It provides the series also for the correlation length and the wave
function renormalization constant. The high order behaviour of
the series coefficients allow for the determination of critical couplings,
including critical exponents.

The effective model has a complex phase structure.
In particular there are both regions in coupling constant space with
mean field and with Ising behaviour.
However, the coupling constants are not free but have been
computed already by dimensional reduction.
They identify the Ising region
as that part which describes the finite temperature Gross-Neveu model.

It would be interesting to supplement these considerations by other means.
In the large $N$ expansion, $N^{-1}$ acts as a real coupling constant.
The parity symmetry of the model does not depend on $N$.
This promises an investigation by means of the renormalization group,
with $N^{-1}$ kept as a running coupling constant.
This investigation should reveal a non-trivial infrared fixed point
at some finite $N^*<\infty$.
For instance, one could try an $\epsilon$-expansion about $1+1$
dimensions.

What do we learn for the QCD chiral phase transition?
The situation is similar to the Gross-Neveu model.
The effective model now is a three-dimensional, pure bosonic
theory. Fermions do not act as dynamical degrees of freedom
on the infrared scale $R\geq T^{-1}$, but they determine the
bosonic interactions.
The $\sigma$-model scenario of the two flavour case is not ruled out,
with critical exponents of the three-dimensional O(4) model.
This is also supported by recent numerical investigations
(\cite{boyd}, cf. also the contribution of Frithjof Karsch
to these proceedings).

Dimensional reduction has been proved to be a powerful method to study
infrared properties of the QCD plasma phase.
The reduction step from four to three dimensions is normally done by
means of renormalized perturbation theory, that is
as an expansion with respect to the renormalized gauge coupling constant
$g_{\rm ren}(R,T)$
at temperature $T$ and length scale $R$.
In this way there is analytic control of the cutoff and volume dependence.
On lattices with moderate extension in the temperature direction,
these effects are rather large.
This implies a considerable complication for numerical investigations.
Their analytic control allows to remove these effects.
In this way it is possible to determine screening masses of the quark gluon
plasma already on small lattices.

This works fine for temperatures at least above twice the phase transition
temperature $T_c$.
Compared to the Gross-Neveu model we have one more spatial dimension.
From dimensional power counting this implies that 
interactions of higher operators are suppressed by inverse powers
of the temperature.
We do not have to rely on a large number of flavours any more
to truncate those terms. However,
approaching the phase transition, the renormalized gauge coupling constant
becomes large, and we can no longer rely on perturbation theory.
The relevant coupling constants of the effective model have to be computed by
non-perturbative means.
Therefore, the universality class of the two-flavour chiral transition
in QCD is still an open question.

%

%
% ---- Bibliography ----
%


\begin{thebibliography}
%
%
\bibitem{}{ambjorn}{Ambj\o rn (1979)}
Ambj\o rn,\, J. (1979):
On the decoupling of massive particles in field theory,
Comm.\, Math.\, Phys. {\bf 67}, 109-119
%
\bibitem{}{appelquist}{Appelquist and Carazonne (1975)}
Appelquist,\, T. and Carazzone,\, J. (1975):
Infrared singularities and massive fields,
Phys.\, Rev. {\bf D11}, 2856-2861
%
\bibitem{}{boyd}{Boyd, Karsch, Laermann and Oevers (1996)}
Boyd,\, G., Karsch.\, F., Laermann,\, E. and Oevers,\, M. (1996):
Two flavour QCD phase transition,
talk given at the 10th International Conference on Problems of
Quantum Field Theory, Alushta, Ukraine, 13-17 May 1996,
Univ. Pisa preprint IFUP-TH-40-96,
and e-Print archive: hep-lat@xxx.lanl.gov 9607046
%
\bibitem{}{butera1}{Butera and Comi (1997)}
Butera,\, P. and Comi,\, M. (1997):
2n-point renormalized coupling constants in the
three-dimensional Ising model:
Estimates by high temperature series to order $\beta^{17}$,
Phys.\, Rev. {\bf E55}, 6391-6396
%
\bibitem{}{campostrini1}{Campostrini, Pelissetto, Rossi and Vicari (1996)}
Campostrini,\, M., Pelissetto,\, A., Rossi,\, P. and Vicari,\, E. (1996):
A strong coupling analysis of two-dimensional O(N) sigma models with
$N\geq 3$ on square, triangular and honeycomb lattices,
Phys.\, Rev.\, {\bf D54}, 1782-1808
%
\bibitem{}{chevalley}{Chevalley (1997)}
Chevalley,\, C. (1997):
{\it The Algebraic Theory of Spinors
and Clifford Algebras} 
(Springer, Berlin, Heidelberg)
%
\bibitem{}{daveiga2}{de Calan, Faria da Veiga, Magnen, Seneor (1991)}
de Calan,\, C., Faria da Veiga,\, P.A., Magnen,\, J., Seneor,\, R. (1991):
Constructing the three-dimensional Gross-Neveu model with a
large number of flavor components.
Phys.\, Rev.\, Lett. {\bf 66}, 3233-3236
%
\bibitem{}{daveiga1}{Faria da Veiga (1991)}
Faria da Veiga,\, P.A., (1991):
PhD thesis, University of Paris (unpublished).
%
\bibitem{}{itzykson}{Itzykson and Drouffe (1989)}
Itzykson,\, C. and Drouffe,\, J.-M. (1989):
{\it Statistical field theory II}
(Cambridge University Press, Cambridge)
%
\bibitem{}{petersson1}{K\"arkk\"ainen, Lacock, Petersson and Reisz (1993)}
K\"arkk\"ainen,\, L., Lacock,\, P., Petersson,\, B. and Reisz,\, T. (1993):
Dimensional Reduction and Colour Screening in QCD,
Nucl.\, Phys. {\bf B395}, 733-746
%
\bibitem{}{kajantie1}{Kajantie, Laine, 
Rummukainen and Shaposhnikov (1996)}
Kajantie,\, K., Laine,\, M., 
Rummukainen,\, K. and Shaposhnikov,\, M. (1996):
The electroweak phase transition: a nonperturbative analysis,
Nucl.\, Phys. {\bf B466}, 189-258
%
\bibitem{}{kapusta}{Kapusta (1989)}
Kapusta,\, J. I. (1989):
{\sl Finite-temperature field theory}
(Cambridge University Press, Cambridge)
%
\bibitem{}{kk1}{Kocic and Kogut (1994)}
Kocic,\, A., Kogut,\,J. (1994):
Can sigma models describe the finite temperature
chiral transitions?,
Phys.\, Rev. \, Lett. {\bf 74}, 3109-3112
%
\bibitem{}{lacock}{Lacock and Reisz (1993)}
Lacock,\, P. and Reisz,\, T. (1993):
Dimensional reduction of QCD and screening masses
in the quark gluon plasma,
Nucl.\, Phys. (Proc. Suppl.) {\bf B30}, 307-314
%
\bibitem{}{landsman}{Landsman (1989)}
Landsman,\, N.P. (1989):
Limitations to dimensional reduction at high temperature,
Nucl.\, Phys. {\bf B322}, 498-530
%
\bibitem{}{luescher2}{L\"uscher (1977)}
L\"uscher,\, M. (1977):
Construction of a selfadjoint, strictly positive
transfer matrix for euclidean lattice gauge theories,
Comm.\, Math.\, Phys. {\bf 54}, 283-292
%
\bibitem{}{luescher0}{L\"uscher and Weisz (1988)}
L\"uscher,\, M. and Weisz,\, P. (1988):
Application of the linked cluster expansion to the $n$-component
$\phi^4$ theory,
Nucl.\, Phys. {\bf B300}, 325-359
%
\bibitem{}{luescher1}{L\"uscher, Sint, Sommer, Weisz and Wolff (1996)}
L\"uscher,\, M., Sint,\, S., Sommer.\, R., Weisz,\, P.
and Wolff,\, U. (1996):
Nonperturbative O(a) improvement of lattice QCD,
Nucl.\, Phys.\, {\bf B491}, 323-343
%
\bibitem{}{hildegard1}{Meyer-Ortmanns (1996)}
Meyer-Ortmanns,\,H. (1996):
Phase Transitions in Quantum Chromodynamics,
Rev.\, Mod.\, Phys. {\bf 68}, 473-598
%
\bibitem{}{hildegard2}{Meyer-Ortmanns and Reisz (1997)}
Meyer-Ortmanns,\,H. and Reisz,\, T. (1997):
Critical phenomena with convergent series expansions
in a finite volume,
Jour.\, Stat.\, Phys. {\bf 87}, 755-798
%
\bibitem{}{osterwalder}{Osterwalder and Schrader (1975)}
Osterwalder,\, K., Schrader,\, R. (1975):
Axioms for Euclidean Green's functions II,
Comm.\, Math.\, Phys. {\bf 42}, 281-305
%
\bibitem{}{park1}{Park, Rosenstein, Warr (1989a)}
Park,\,S., Rosenstein,\, B., Warr,\, B. (1989a):
Four-fermion theory is renormalizable in 2+1 dimensions.
Phys.\, Rev.\, Lett. {\bf 62}, 1433-1436
%
\bibitem{}{park2}{Park, Rosenstein, Warr (1989b)}
Park,\,S., Rosenstein,\, B., Warr,\, B. (1989b):
Thermodynamics of (2+1) dimensional four-fermion models.
Phys.\, Rev. {\bf D39}, 3088-3092
%
\bibitem{}{pisarski}{Pisarski and Wilczek (1984)}
Pisarski,\, R.\, D. and Wilczek,\, F. (1984):
Remarks on the chiral phase transition in chromodynamics.
Phys.\, Rev. {\bf D29}, 338-341
%
\bibitem{}{pordt1}{Pinn, Pordt and Wieczerkowski (1994)}
Pinn,\, K., Pordt,\, A. and Wieczerkowski,\, C. (1994):
Computation of hierarchical renormalization group fixed points
and their $\epsilon$-expansion,
Jour.\, Stat.\, Phys. {\bf 77}, 977-1005
%
\bibitem{}{pordt2}{Pordt (1996)}
Pordt,\, A. (1996):
A convergence proof for linked cluster expansions,
Univ. M\"unster preprint MS-TPI-96-05,
and e-Print archive: hep-lat@xxx.lanl.gov 9604010
%
\bibitem{}{pordt3}{Pordt  and Reisz (1996)}
Pordt,\, A. and Reisz,\, T. (1996):
Linked cluster expansions beyond nearest neighbour interactions:
Convergence and graph classes (1996):
Univ. Heidelberg preprint HD-THEP-96-09 and
Univ. M\"unster preprint MS-TPI-96-06,
and e-Print archive: hep-lat@xxx.lanl.gov 9604021,
to appear in Int. J. Mod. Phys. A
%
\bibitem{}{porteous}{Porteous (1995)}
Porteous,\, I. R. (1995):
{\sl Clifford algebras and the classical groups}
(Cambridge University Press, Cambridge)
%
\bibitem{}{rajagopal}{Rajagopal and Wilczek (1993)}
Rajagopal,\, K. and Wilczek,\, F. (1993):
Static and dynamic critical phenomena of a second order
QCD phase transition,
Nucl.\, Phys.\, {\bf B399}, 395-425
%
\bibitem{}{reisz0}{Reisz (1988)}
Reisz,\, T. (1988):
A power counting theorem for Feynman integrals on the lattice,
Comm.\, Math.\, Phys. {\bf 116}, 81-126, and:
Renormalization of Feynman integrals on the lattice,
Comm.\, Math.\, Phys. {\bf 117}, 79-108
%
\bibitem{}{reisz1}{Reisz (1992)}
Reisz,\, T. (1992):
Realization of dimensional reduction at high temperature,
Z.\, Phys. {\bf C53}, 169-176
%
\bibitem{}{reisz2}{Reisz (1995a)}
Reisz,\, T. (1995a):
Advanced linked cluster expansion:
Scalar fields at finite temperature,
Nucl.\, Phys. {\bf B450}[FS], 569-602
%
\bibitem{}{reisz3}{Reisz (1995b)}
Reisz,\, T. (1995b):
High temperature critical O(N) field models by LCE series,
Phys.\, Lett. {\bf B360}, 77-82
%
\bibitem{}{rothe}{Rothe (1997)}
Rothe,\, H. J. (1997):
{\sl Lattice gauge theories, an introduction}, 2nd edition
(World Scientific, Singapore)
%
\bibitem{}{wilczek}{Wilczek (1992)}
Wilczek,\, F. (1992):
Application of the renormalization group to a second-order
QCD phase transition.
Int.\, Jour.\, Mod.\, Phys.\, {\bf A7}, 3911-3925
%
\bibitem{}{wortis}{Wortis (1974)}
Wortis,\, M. (1974):
Linked cluster expansion, in {\sl Phase Transitions and Critical
Phenomena} Vol. 3,
Domb,\, C. and Green,\, M. S., eds. (Academic Press, London)
%
\bibitem{}{zinoviev}{Zinoviev (1995)}
Zinoviev, \, Yu.\, M. (1995):
Equivalence of Euclidean and Wightman field theories,
Comm.\, Math.\, Phys. {\bf 174}, 1-28
%
\end{thebibliography}
\end{document}